\documentclass{aa}

\usepackage{amssymb}                                            
\usepackage{amsmath}                                            
\usepackage{graphicx}                                          
\usepackage[usenames]{color}

\usepackage{txfonts}
\usepackage{natbib}

\defcitealias{lip-sha2002}{LS02}
\defcitealias{lip-sha2000}{LS00}
\begin{document}

\voffset=-1.0cm

\newcommand{\be}{\begin{equation}}
\newcommand{\ee}{\end{equation}}
\newcommand{\bea}{\begin{eqnarray}}
\newcommand{\eea}{\end{eqnarray}}
\newcommand{\mon}{A\,0620$-$00\protect}
\newcommand{\mus}{GRS\,1124$-$68\protect}
\newcommand{\Mbh}{M_{\mathrm{bh}}}
\newcommand{\Ledd}{L_{\mathrm{Edd}}}
\newcommand{\Rin}{R_{\mathrm{in}}}
\newcommand{\ax}{a_{\mathrm{X}}}
\newcommand{\Fx}{F_{\mathrm{X}}}
\newcommand{\alert}{\color{red}}
\newcommand{\changed}{\color{blue}}

\newcommand{\att}[1]{\marginpar[$>->->$]{$<-<-<$}{\uline{#1}}}
\newcommand{\out}[1]{\marginpar[$>->->$]{$<-<-<$}{\sout{#1}}}
\newcommand{\dd}{^\mathrm{d}}

\title{
Modeling of non-stationary accretion disks in X-ray novae
A\,0620$-$00 and GRS\,1124$-$68 during outburst
}

\author{
V.\,F.~Suleimanov\inst{1,2},
G.\,V.~Lipunova\inst{3},
N.\,I.~Shakura\inst{3}}

\offprints{V.~Suleimanov}
\mail{e-mail: suleimanov@astro.uni-tuebingen.de}

\institute{
Institute for Astronomy and Astrophysics, Kepler Center for Astro and
Particle Physics,
 Eberhard Karls University, Sand 1,
 72076 T\"ubingen, Germany
\and
Kazan State University, Kremlevskaja str., 18, Kazan 420008, Russia
        \and
	    Sternberg Astronomical Institute,
	    Moscow State University, Universitetskij pr., 13, 119992,
	    Moscow, Russia\\
            \email{galja@sai.msu.ru, shakura@sai.msu.ru}
}

\date{Received xxx / Accepted xxx}

   \authorrunning{Suleimanov, Lipunova \& Shakura}
   \titlerunning{Modeling of non-stationary accretion disks in X-ray novae}

\abstract
{}
{We address the task of modeling soft X-ray and optical light
curves of X-ray novae in the high/soft state.}
{The analytic model of viscous
evolution of an externally truncated accretion $\alpha$-disk is used.
Relativistic effects near a Kerr black hole and self-irradiation of an
accretion disk are taken into account.
}
{The model is applied to the
outbursts of X-ray nova Monocerotis 1975 (\object{A\,0620$-$00}) and
X-ray nova Muscae 1991 (\object{GRS\,1124$-$68}). Comparison of
observational data with the model yields constraints on the
angular momentum (the Kerr parameter) of the black holes in A\,0620$-$00
and GRS\,1124$-$68: $0.3-0.6$ and $\leq 0.4$, and on the viscosity
parameter $\alpha$ of the disks: $0.7-0.95$ and $0.55-0.75$.
We also conclude that the  accretion disks should have an
effective geometrical thickness $1.5-2$ times greater than
the theoretical value of the distance between the photometric
layers.}
{}


\maketitle
%

\section{Introduction\label{s.intro}}

X-ray novae are the low-mass X-ray close binary systems~(LMXBs) with a
relativistic component, a black hole, or a neutron star~\citep[see,
e.g.,][]{tan-shi1996,esin_et2000, cherep2000}. Much attention has been
drawn to X-ray novae because they provide black hole candidates with the
most reliable mass determinations~\citep{cherep2000,mcc-rem2003}. The
secondary component is a red dwarf, which nearly fills its Roche lobe,
and transfer rates are estimated to be as low as {$10^{-10}-10^{-12}$}
$M_{\sun}$/yr~\citep{tan-shi1996}. Once every several decades, the
accretion rate onto the central object rises (up to $10^{-8}
M_{\sun}$/yr), and the system flares up as a Nova at all wavelengths.

In a significant number of X-ray novae the X-ray flux declines in an
exponential fashion after the peak of an outburst, with a characteristic
time of $30-40$~days~\citep[see, e.g.,][]{mcc-rem2003}. The optical flux
also decays exponentially but about two times more slowly. The X-ray
emission of the majority of X-ray novae during an outburst can be
interpreted as the sum of a hard power-law component and a soft
component produced by a multi-temperature black body disk with a maximum
temperature \hbox{$0.5 - 1$~keV}~\citep{tan-shi1996}. As a rule, during
the first several tens of days after the peak, the soft component
dominates the power-law component (this stage is called the
high/soft state). After 100--150~days the spectrum becomes harder as the
power-law component prevails. An outburst usually lasts for several
months.

Up to now the mechanism of the outbursts in X-ray novae has not been
completely understood, but it most probably involves an instability of the
outer disk regions~\citep{kin-rit1998,cherep2000}. This instability,
related to hydrogen ionization, was initially proposed to explain dwarf
novae, in which accretion occurs onto a white
dwarf~\citep{meye-meye1981,smak1984a, cannizzo1993}. Subsequently, the
concept was extended to X-ray novae. Van Paradijs (1996) followed by
\cite{kin-rit1998}  suggested that irradiation of the outer parts of the
accretion disk by central X-rays has a significant impact on the
dynamics of the evolutionary cycle in X-ray novae. Such
irradiation is low in the dwarf novae \citep{dubus_et2001,lasota2001}.

To explain the spectral evolution of X-ray novae, a number of models
have been proposed~\citep[see review by][]{mcc-rem2003}.
\citet{esin_et1997} proposed a scenario involving a change of  cooling
mechanisms with decreasing accretion rate. According to their model, at
the outburst peak and through the subsequent 100--150~days, the disk is
geometrically thin and optically thick and embedded in a hot corona. As
the accretion rate falls, the disk, begining from the center, gradually
becomes advection-dominated, geometrically thick, and optically
thin~\citep{narayan_et1998}. Using a succession of stationary models
with different accretion rates, \citet{esin_et1997, esin_et2000}
describe the X-ray and optical light curves and the spectral evolution
of X-ray nova Muscae 1991 and X-ray nova Monocerotis 1975. To explain
the observed X-ray-to-optical flux ratio they considered an irradiation
of the disks and assumed that 80--90\% of the incident X-ray flux is
thermalized in the disks, which are thicker than given by the standard
theory.

Naturally, time-dependent phenomena in disks call for a dynamical model.
Whereas a stationary hydrostatic approximation can be kept to describe the
vertical structure of the disk, a non-stationary model should primarily
address its radial structure.

A theory of time-dependent accretion $\alpha$-disks in binary systems
has been worked out by \citet{lip-sha2000} (hereafter
\citetalias{lip-sha2000}) and subsequently used to describe X-ray and
optical light curves of X-ray nova Monocerotis 1975 and X-ray nova
Muscae 1991~\citep[][hereafter \citetalias{lip-sha2002}]{lip-sha2002}.
The model applies to the first several tens of days after the outburst
peak, when the spectral state is high/soft, the hydrogen is ionized
in the whole disk, and before the secondary maximum. The disk is considered
to be externally truncated due to tidal interactions in a binary.
Fitting the observational light curves of the two X-ray novae provided
estimates of the $\alpha$-parameter and relationships between
the black hole mass and the distance to the systems. Actually, a new method of
estimating the $\alpha$-parameter of accretion disks was proposed, and the
$\alpha$-parameter turned out to be rather high, 0.2--0.6.

In the present work, we develop the model of
\citetalias{lip-sha2002} further. We include the effect of self-irradiation of
the disk and take into account the general relativistic effects on the
radial structure of the disk and on light propagation near a Kerr black
hole. The last effect increases the degree of irradiation remarkably
compared to the plain space-time geometry, even for a non-rotating black
hole~\citep{suleimanov_et2007e}.

In Sect.~\ref{s.model} theoretical aspects of the model are considered.
The modeling procedure and general features of the solution are
described in Sect.~\ref{s.modeling}. Application of the model to the
X-ray novae \mon{} and \mus{} follows in Sect.~\ref{s.results}.

\section{The model\label{s.model}}
\subsection{Time-dependent accretion disk\label{ss.disk}}
The theory of time-dependent accretion has been put forward in the
work of \citet{lyub-shak1987} and developed for an $\alpha$-disk in a
close binary system by \citetalias{ lip-sha2000}.

A diffusion-type differential equation of non-stationary accretion is
solved analytically assuming specific outer boundary conditions.The
size of the disk is fixed due to tidal effects, and the accretion rate
through the outer boundary is negligible compared with the rate through the
inner boundary of the disk. The opacity coefficient is a power-law
function of density $\rho$ and temperature $T$. Separate regimes with
scattering and Kramers absorption are considered. The vertical structure
is treated anaytically on the basis of a method proposed
by~\citet{ket-sha1998}. As a result, the accretion rate at each radius
is a power-law function of time and of some  other parameters. In the case of
Kramers opacity, it is given by~\citepalias{lip-sha2000}:
\begin{equation}
 \dot M(R,t) = 2 \,\pi\, \frac{1.224\,
y(\xi)}{h_0} \left (
\frac{h_0^{14/5}}{D\,(t+\delta t)} \right )^{10/3}\, .
\label{Mdot_sol}
\end{equation}
Here $h_0=\sqrt{GM_\mathrm{bh}R_\mathrm{out}}$ is the specific angular
momentum of the disk matter at the outer radius; $\delta t$ has
dimensions of  seconds  and normalizes the initial accretion rate; $D$ is a
dimensional ``diffusion constant'' depending on the $\alpha$-parameter,
the black hole mass, and the opacity coefficient $\kappa_0$; function
$y(\xi)$ is a polynomial truncated to three terms:
\be
y(\xi)=1.43 - 1.61\, \xi^{5/2} + 0.18\, \xi^{5},
\label{y(x)}
\ee
where $\xi = \sqrt{R/R_{\rm out}}$ is the ratio of the angular momentum
at a given disk radius to the angular momentum at the outer disk radius.

One can infer from~(\ref{Mdot_sol}) that the accretion rate near the
central object hardly depends on $R$. The disk dynamics (variations in
the accretion rate, in particular) are defined by its structure at outer
radii, because it is there that the characteristic viscous timescale is
the longest. Changes in the accretion rate, occurring on a smaller time
scale at inner radii, respond to the course of evolution at the outer
disk. Thus, to analyze the light curves of a fully ionized accretion
disk in any spectral band, one can use solution~(\ref{Mdot_sol}) written
for the case of Kramers opacity $\kappa = \kappa_0 \rho T^{-7/2}$.

We take $\kappa_0 = 5 \times
10^{24}~\mbox{cm}^5\,\mbox{K}^{7/2}\,\mbox{g}^{-2}$~\citep[][chap.
5]{fkr_book2002}, which is  the best fit to the real opacity of matter
with solar abundances when hydrogen is fully ionized and bound-free
absorption by heavy elements is present. Then the diffusion constant $D$
takes the form
\be
D= 7.56 \times 10^{34} \alpha^{4/5} (\mu/0.5)^{-3/4}
\frac{M_\mathrm{bh}}{M_{\odot}}\, .
\label{D}
\ee
The numerical constant in~(\ref{D}) incorporates the opacity coefficient
$\kappa_0$ and a factor about $1$ that accounts for the
vertical structure solution~\citepalias[for more details
see][]{lip-sha2000,lip-sha2002}.

\begin{figure}
\begin{center}
\resizebox{\hsize}{!}{
   \includegraphics[trim= 1cm 1cm 1cm 1.5cm,clip]{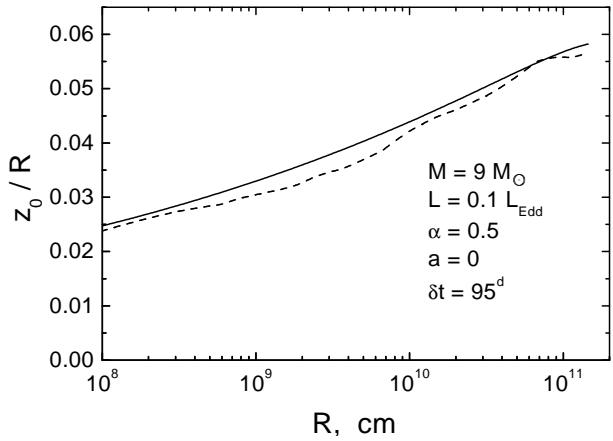}}
\end{center}
\caption{
Relative half-thickness of the disk calculated for Kramers opacity
 (solid line) and for the tabulated values of opacity~(dashed line).
  The pertubations in the dashed line are caused by opacity variations.
\label{z_R.ris}
}
\end{figure}

The key parameters for computing the emission of the disk are the effective
temperature and the half-thickness of the disk, crucial for calculating
self-irradiation. \citetalias{lip-sha2000} provides the half-thickness
of the disk for the zones with Kramers opacity:
\begin{multline}
 \frac{z_0(R)}{R} = 0.135 \alpha^{-1/2}
\left(\frac{M_\mathrm{bh}}{M_{\odot}}\right )^{-1/4}\,
\left(\frac{R}{R_\mathrm{out}} \right )^{1/20} \times \\ \times
 f(\xi)^{3/20} \,\left (
 \frac{R_\mathrm{out}}{R_{\odot}} \right )^{3/4} \left (\frac{t+\delta
 t}{10^d} \right )^{-1/2}\, .
\label{eq.z/r}
\end{multline}
Here the numerical constant contains a factor about $2.6$ that
accounts for the vertical structure solution and thus depends on the
value $\kappa_0$ (see \citet{suleimanov_et2007e} and Eq.~(36)
therein). Function $f(\xi)$ is such that $y(\xi)=f'(\xi)$ and
\be
f(\xi)=1.43\,\xi - 0.46\, \xi^{7/2} +0.03\,\xi^6\, .
\ee
Figure~\ref{z_R.ris} shows the ratio $z_0/R$ versus radius obtained by
solving the vertical structure equations with the tabulated opacity
values calculated for the solar chemical
abundance~\citep{kurucz1994_n19-22,suleim1991,lip-sul2004}. It can be
seen that the half-thickness of the disk calculated from the real
(tabulated) opacity is reproduced satisfactorily by
Eq.~\eqref{eq.z/r}.

The effective temperature of the disk may be written as
\be
\sigma T^4_\mathrm{vis}(R,t) = \frac{3}{8\pi}\, \dot M(0,t)\,
\omega_{\mathrm{K}}^2(R)\, \Phi(R)\, ,
\label{Teff}
\ee
where $\sigma$ is the Stephan-Boltzmann constant;
$\omega_{\mathrm{K}}=\sqrt{GM/R^3}$; $\Phi(R)$ is a ``structure function''
such that for a stationary, not-truncated disk in the Newtonian limit
$\Phi(R)=1-\sqrt{\Rin/R}$. Taking general relativistic effects into account
following \citet{page-thor1974} and \citet{rif-her1995} with zero torque
at the inner disk radius, we have
\begin{multline}
\Phi(R) =  \left[ \left( \frac{R}{R_0}\right) ^3 \,
\frac{x-x_0-1.5\,a\ln\frac{x}{x_0}-A-B-C}{x^4(x^3-3x+2a)} \right] \\
\times \left[ \sqrt{\frac{R_\mathrm{out}}{R}}\,
\frac{ f(\xi)}{y(0)}\right] \, ,
\label{eq.struct_func}
\end{multline}
where
\bea
    A&=&\frac{3(x_1-a)^2}{x_1(x_1-x_2)(x_1-x_3)} \ln\frac{x-x_1}{x_0-x_1}\, ,\nonumber\\
    B&=&\frac{3(x_2-a)^2}{x_2(x_2-x_1)(x_2-x_3)} \ln\frac{x-x_2}{x_0-x_2}\, ,\nonumber\\
    C&=&\frac{3(x_3-a)^2}{x_3(x_3-x_1)(x_3-x_2)} \ln\frac{x-x_3}{x_0-x_3}\, ,\\
   x_1&=&2\cos\left(\frac{\arccos{a}-\pi}{3}\right),\quad
    x_2 = 2\cos\left(\frac{\arccos{a}+\pi}{3}\right), \nonumber\\
    x_3&=&-2\cos\left(\frac{\arccos{a}}{3}\right),\quad
          x_0=\sqrt{\frac{\Rin}{R_0}},\quad
	                      x=\sqrt{\frac{R}{R_0}},\nonumber
\eea
and $R_0=G\,M_\mathrm{bh}/c^2$ . The term in the first square brackets
of Eq.~\eqref{eq.struct_func} represents a general relativistic correction
and approaches 1 at large radii. The term in the second square brackets
of Eq.~\eqref{eq.struct_func} results from the non-stationary solution, and
it equals approximately $1/1.43$ near the outer radius and $\approx 1 $
at small radii.

 At the outer edge of the disk there are additional sources of heating. Some type of
collision between the disk and the incoming matter may cause a development of a hot area.
Moreover, the tidal interactions rise steeply at $R_\mathrm{out}$~\citep{ich-osa1994}
 and lead to a corresponding energy release.
 We believe that neither of the above effects
contributes significantly to the optical flux during the outburst and do
not include them in the present study.

\subsection{X-ray spectrum\label{ss.x-spectrum}}

In the vicinity of a compact relativistic object, the outgoing X-ray
spectrum is disturbed by the effects of Doppler boosting due to the disk
rotation, gravitational focusing, and the gravitational redshift
\citep{cunnin1975}. To calculate the spectrum of a disk, as detected by
an observer at infinity, we use a computer code by
\citet{speith_et1995}, which provides a transfer function allowing for
these effects.

The transfer function is applied to the local X-ray spectra, which are
diluted Planck spectra with a color temperature $T_\mathrm{c}(R) =
f_\mathrm{c}\, T_\mathrm{vis}(R)$ due to Compton scattering, where
$f_\mathrm{c}=1.7\pm0.2$~\citep{shi-tak1995,davis_et2005}.

\begin{figure}
  \begin{center}
  \resizebox{\hsize}{!}{\includegraphics[trim= 0cm 0.5cm 0cm 1.5cm,clip]
     {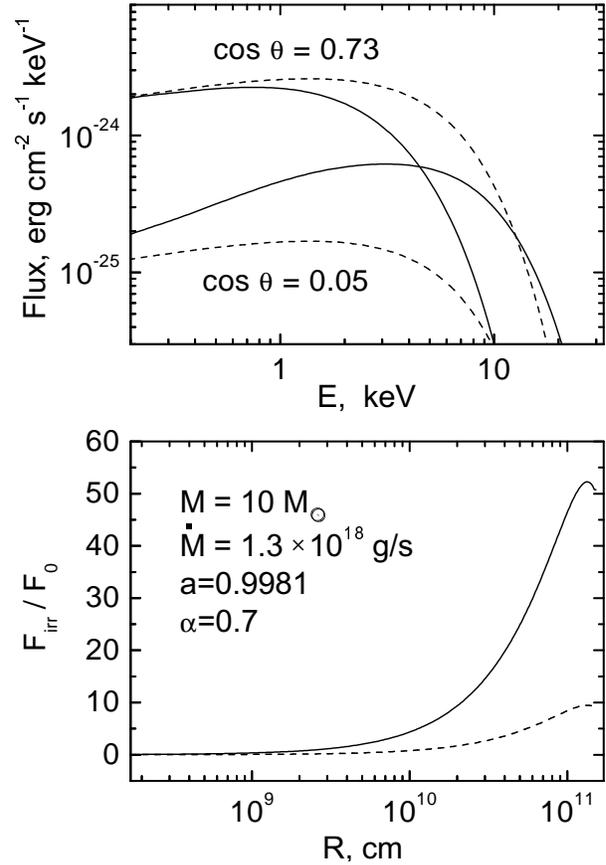}}
  \end{center}
  \caption{
  Upper panel: X-ray spectra  of the accretion disk
around a Kerr black hole calculated for different disk
inclinations. Lower panel:  incident-to-intrinsic flux ratio vs. radius.
For both panels the Kerr black hole has  $a=0.998$.
The solid lines show the impact
of the relativistic effects on light propagation and can be compared with
results obtained in the Newtonian limit shown by the dashed lines.
  \label{oto.ris}
  }
\end{figure}
Figure~\ref{oto.ris} illustrates the general relativistic effects on the
spectrum of the accretion disk around a Kerr black hole (the Kerr
parameter $a=0.998$). Shown are the spectrum as seen by a remote
observer for a disk inclination $\theta = 43\degr$ ($\cos \theta=0.73$)
and the spectrum for $\theta = 87\degr$ ($\cos \theta=0.05$). Naturally,
the spectral shape of the X-ray emission received by the outer disk
($\theta$ is large) differs significantly from what is seen by a remote
observer. The intensity of the illuminating flux is significantly higher
than calculated without general relativistic effects.

There are different methods of computing general relativistic effects on
light propagation. An example of another implementation is one by
\cite{li_et2005}, which is realized in the model ``kerrbb'' in the xspec
package. However, kerrbb is restricted to inclination angles $\leq
85^\mathrm{o}$. Thus we cannot readily use it to obtain the flux
illuminating the outer disk, the calculation of which we describe in the
following section.

To calculate the X-ray flux at the Earth we take 
interstellar absorption into account~\citep{mor-mcc1983}.

\subsection{Self-irradiation of the disk\label{ss.self-irradiation}}

\begin{figure}
  \begin{center}
  \resizebox{\hsize}{!}{\includegraphics[trim= 0cm 2.8cm 0cm 2cm,clip]{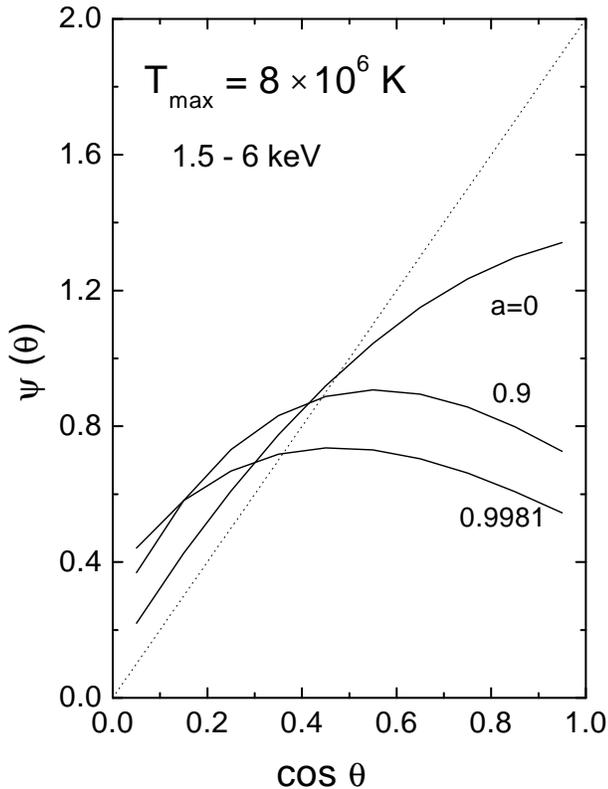}}
  \end{center}
  \caption{
  Function $\Psi(\theta)$  for the different values of the Kerr parameter.
See also Fig.~9 in~\citet{suleimanov_et2007e}.
  \label{fig.psi}
  }
\end{figure}

The flux falling onto a unit area that is normal to the emission
outgoing at an angle $\theta$ to the disk axis, can be written as
\be
 F(r,\theta)
   =  \frac{L_\mathrm{X}}{4\, \pi\, r^2}\, \Psi(\theta)\, ,
   \label{irrad.flux}
\ee
where $L_\mathrm{X}$ is the luminosity of the central X-ray source and
$r$ the distance of the unit area from the center. In plain
space-time geometry, for a flat disk as the central source,
$\Psi(\theta) = 2\,\cos(\theta)$. We use the function $\Psi(\theta)$
provided by the code of \citet{speith_et1995}. Examples of
$\Psi(\theta)$ for some values of the Kerr parameter are shown in
Fig.~\ref{fig.psi}

To calculate flux absorbed by a unit area of the disk, one should allow
for not all of the incident flux being
thermalized~\citep{sha-sun1973,vanparadijs1996,kin-rit1998}:
\be
   F_\mathrm{irr}(R)= \eta\, \frac{L_\mathrm{X}}{4 \pi R^2}\,  \Psi(\theta)
   \,\left [\frac{dz_0}{d R} - \frac{z_0}{R} \right ] ,
\label{eq.F_irr}
\ee
where $\eta$ is a thermalization coefficient, and the term in square
brackets stands for the inclination of the disk surface to the incident
rays. Assuming the disk to be geometrically thin, the distance $r$ is
replaced by the radius of the disk $R$ at which the area in question is
located.

The geometrical factor in the square brackets of Eq.~\eqref{eq.F_irr} is
remarkably different for the cases of the time-dependent model and a
stationary disk. A stationary disk with $\dot M = \dot M(0,t)$ has
approximately the same $z_0/R$ as in Eq.~\eqref{eq.z/r}. However, value
$\mathrm{d}z_0/\mathrm{d}R$ differs significantly. Using
Eq.~\eqref{eq.z/r} we obtain the geometrical factor for a
non-stationary, externally-truncated disk,
\be
  \frac{dz_0}{d R} - \frac{z_0}{R}  =
 \frac{z_0}{R}\left(\frac{1}{20} + \frac{3}{40} \sqrt{\frac{R}{R_{\rm out}}}
    \frac{y(\xi)}{f(\xi)} \right)~,
\label{eq.geom}
\ee
which is $\approx 1/20\,(z_0/R)$ for $R\approx R_{\rm out}$, while the
geometrical factor for a steady-state disk model (Shakura \& Sunyaev
1973) is $1/8\,(z_0/R)$; thus, the time-dependent disk near the outer
radius is less irradiated.

The factor $\eta$ can be thought of as a coefficient for the reprocessing of
incident X-rays (``reprocessing efficiency''), i.e. the fraction of
X-rays absorbed in the under-photosphere layers. Values of $\eta$
suggested by researchers differ significantly. For low-mass X-ray
binaries, \citet{dejong_et1996} obtained a low value for reprocessing
efficiency $\eta \sim 0.1$. In contrast, mo\-de\-ling the outbursts of
\mus{} and \mon{}, \citet{esin_et2000} assume that almost all the
incident X-ray flux is thermalized, $\eta =0.8-0.9$.

Taking another approach, a value for the reprocessing efficiency can be
derived from an accurate consideration of the atmosphere of an accretion
disk exposed to the incident radiation~\citep{suleim_et1999}. For a disk
with ionized hydrogen, a decisive factor is the spectrum of the incident
X-rays. Only rather hard X-rays, of energy $E>2$~keV, penetrate  the
layers where the optical continuum is produced. Softer X-rays are
absorbed higher in the atmosphere and reprocessed to the far
UV~\citep{suleim_et1999,elk-wic1999}. To raise the optical emission from
the disk, the temperature should increase at the depths where the
optical continuum is formed. However, most of the X-rays with energies
over 10~keV are reflected due to electron scattering. Thus, it is
mainly X-rays in the interval $2-10$~keV that are reprocessed to optical
emission. The high/soft spectral state of X-ray novae is characterized
by soft X-ray spectra with a corresponding black body temperature $\sim
0.5 - 1.0$~keV. Using Eq.~(22) from \citet{suleim_et1999} we obtain
low values of the reprocessing efficiency, $0.1-0.2$.

\citet{esin_et2000} generally conclude that the outer parts of
accretion disks in the high/soft state should be significantly thicker
than follows from the standard model of~\citet{sha-sun1973}. Concerning
this result we bring up the following argument. The irradiation cannot
increase the thickness of the disk beyond the values predicted by
Eq.~\eqref{eq.z/r}, either in the regions where Kramers opacity applies,
or in the regions with partly ionized
hydrogen~\citep{dubus_et1999,suleimanov_et2007e}. However,
a layer of hot coronal gas  could be present above the disk due to the irradiation
of the outer disk by hard X-rays~\citep{begelman_et1983,
begel-mckee1983, ko-kallman1994,
rozan-czern1996,proga-kallman2002,jimenez_et2002}. Such a corona,
optically thick in the radial direction, effectively increases the
geometrical thickness of the disk and augments the degree of
self-irradiation. Furthermore, the corona could be inhomogeneuos and
contain optically thick gas blobs as proposed for supersoft X-ray
sources by \citet{suleim_et2003}. In this case the reprocessing
efficiency can increase significantly due to multiple scattering between
the blobs.

There is another possibility for increasing a disk's thickness.
Recent simulations of magnetorotational turbulence
\citep[e.g.][]{Miller_St2000,tur_04,hir2006} show that magnetic pressure is
significant in the upper layers of the disk, probably resulting in a
vertically extended atmosphere. Thus, the optical-depth unity surface can
be located a factor of two above the ``standard" disk surface
\citep{hir2006}.

In view of the present uncertainty, we consider the reprocessing
efficiency as a parameter. For the same reason, we do not calculate the
half-thickness of the disk using the tabulated values of opacity
coefficient but utilize Kramers approximation. In Eq.~\eqref{eq.F_irr},
the presence of the corona is treated as the increased thickness of the
disk, which equals $z_0$ multiplied by a chosen factor.

 It should be noted that consideration of the hard spectral 
component, which is emitted by a central source (possible central corona) 
and characterized by $\Psi(\theta)\approx 1$, would enhance the
self-irradiation of the disk.

\subsection{Optical emission from the disk\label{ss.optics}}
The spectrum of the time-dependent disk at any moment of time
is calculated by integrating local spectra over radius:
\be
L_{\lambda} = 4\, \pi^2\, \int_{R_\mathrm{in}}^{R_\mathrm{out}}
B_{\lambda}(T)\, R\, \mathrm{d}R\, .
\ee
At the outer parts of the disk the local spectra are
 Planck functions with an effective temperature $T$ increased
due to the intercepted X-ray flux
\be
\sigma\, T^4 = \sigma\, T_\mathrm{vis}^4 + F_\mathrm{irr},
\ee
where $T_\mathrm{vis}$ is the effective temperature~(Eq.~\ref{Teff}) in
the absence of X-ray irradiation. The spectral flux density at the Earth
is given by
\be
f_{\lambda} =
\frac{L_{\lambda}}{2 \pi d^2}\, \cos i\ \exp(-\tau_{\lambda})\, ,
\ee
where $d$  the distance to the system, $i$  the inclination of the
disk, $\tau_{\lambda}$ is the optical depth for interstellar absorption.

\begin{figure}
  \begin{center}
  \resizebox{\hsize}{!}{\includegraphics[trim= 0cm 2.5cm 0cm 2cm,clip]
       {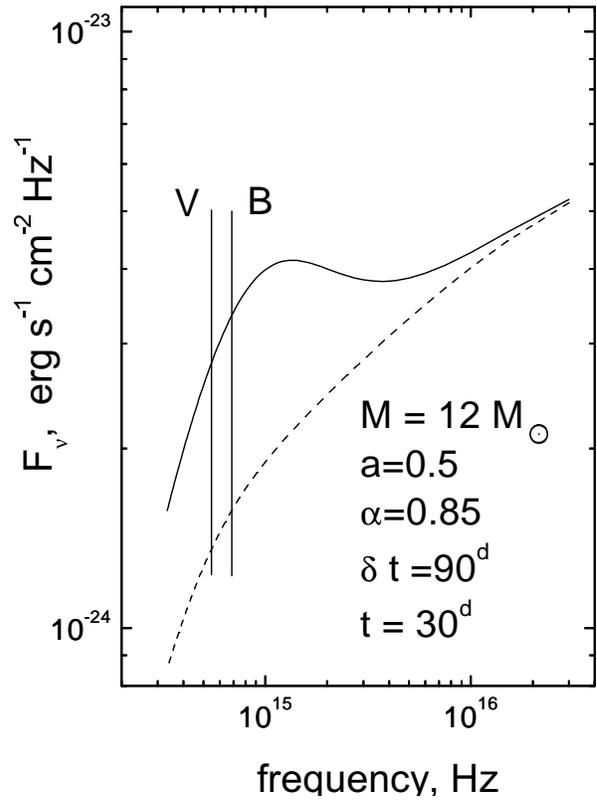}}
  \end{center}
  \caption{
  Optical and UV spectrum of a modeled disk of \mon{} with
self-irradiation (the solid line) and without (the dashed line).
Vertical lines denote the positions of the centers of  $B$ and $V$
bands.
  \label{fig.o_spec}
  }
\end{figure}

The unabsorbed optical and UV spectrum of a self-irradiated disk is
shown in Fig.~\ref{fig.o_spec}.

\section{Modeling procedure\label{s.modeling}}
The physical parameters of X-ray novae (mass, Kerr parameter, viscosity
parameter $\alpha$, etc.) are sought by comparing modeled and observed
light curves. A particular model of the disk is defined by a set of
input physical parameters. Some of these can be varied and some are
assumed to be known from observations.

To compare the model with observations, we chose several values
calculated in a specific way from observational data-- ``fitted light
curve parameters''-- and fit them by the model.

\subsection{Fitted light-curve parameters\label{ss.modeled_p}}

As shown in~\citetalias{lip-sha2002}, the X-ray and optical light
curves of the X-ray novae considered are represented well by linear
functions in ``log(flux)--time'' coordinates throughout the first
40--60~days after the peak. Thus each light curve can be described by a
pair of parameters: the slope and the flux at a reference time, which
are to be fitted. This approach allows one to significantly reduce the
computer time required for finding satisfactory model parameters.

In the course of the modeling, we reproduce the following values: the
slope of the X-ray light curve $\ax$, the slope of the optical light
curve in the $B$ band $a_B$, the X-ray flux $\Fx$, the flux $F_B$, the
color index $B-V$, and, optionally, $U-B$. All parameters, except for the
first two, are calculated for a reference time.

 \subsection{Input parameters of the model\label{ss.input_p}}

The free input parameters are:
 the black hole mass $M_\mathrm{bh}$,
 its Kerr parameter $a$,
 the viscosity parameter $\alpha$ of the disk,
 the time parameter $\delta t$ to fix the accretion rate at time $t=0$,
 and the reprocessing efficiency $\eta$.
Other variable parameters of the model are the color-to-effective
temperature ratio $f_\mathrm{c}$ and  the ratio $z/z_0$, where $z$ is
the height of X-rays' interception.

 The parameters that are more or less fixed include:
 the optical component mass $M_\mathrm{o}$,
 the mass function $f(M)$,
 the period of the binary $P$,
 and the interstellar absorption, expressed by
 parameter $N_\mathrm{H}$ for an  X-ray band
 (the number of H atoms along the line of sight
 per cm$^2$) or by the color excess $E(B-V)$ for optical bands.

The mass of the optical component $M_\mathrm{o}$ and the interstellar
absorption are not very precisely known. In \citetalias{lip-sha2002}
they were varied within narrow limits, and it was shown that the effect
of the variation was small, so  we fix their values.

Given the binary mass, one obtains the semi major axis of the binary
system
\be
\tilde{a} =\left ( \frac{G\, (M_\mathrm{bh}+M_\mathrm{o})\, P^2}{4\,
 \pi^2} \right )^{1/3}\, \label{semiaxis}
\ee
and can calculate the outer radius of the disk approximately. We take
this equal to 80\% of the radius of a sphere whose volume equals that
of the black hole's Roche lobe~\citep{egglton1983}:
\be
  R_\mathrm{out} = 0.8\, \tilde{a}\,
  \frac{0.49\, q^{2/3}}{0.6\, q^{2/3}+\ln(1+q^{1/3})}\, ,
  \label{rout}
\ee
where $q=M_\mathrm{bh}/M_\mathrm{o}$. This approximation agrees with
calculations by \citet{paczynski1977} and with the direct estimates of
disk sizes in cataclysmic variables~\citep{sulkanen_et1981}. One can
also derive the inclination $i$ of the binary plane:
\be
i = \arcsin
\left ( \frac{f(M)\,(M_\mathrm{bh}+M_\mathrm{o})^2}{M_\mathrm{bh}^3}
\right)^{1/3}.
\ee
 It is necessary to mention that the inner disk can have another
inclination to the line of sight. The black hole angular momentum can be
misaligned with that of a binary due to asymmetric previous
supernova explosion. In this case the inner disk will tilt toward to
equatorial plane of the black hole
\citep{bp1975}. We do not consider this possibility in the present study
and assume that the inner and the outer  parts of the disk have the same inclination;
therefore, some of our results can be biased, black hole
angular momentum in particular.

The distance to the system is defined in the course of the modeling. As
the limits on black hole masses and distances are known from
observations, they are used to restrict the allowed ranges of other
parameters.

\subsection{Qualitative investigation of nonstationary accretion in
$\alpha$-disks
\label{qualitative.ss}}

Before proceeding to the model results, we discuss the qualitative
features of the nonstationary model.

\begin{figure}
  \begin{center}
  \resizebox{\hsize}{!}{\includegraphics[trim= 0cm 2.5cm 0cm 2cm,clip]{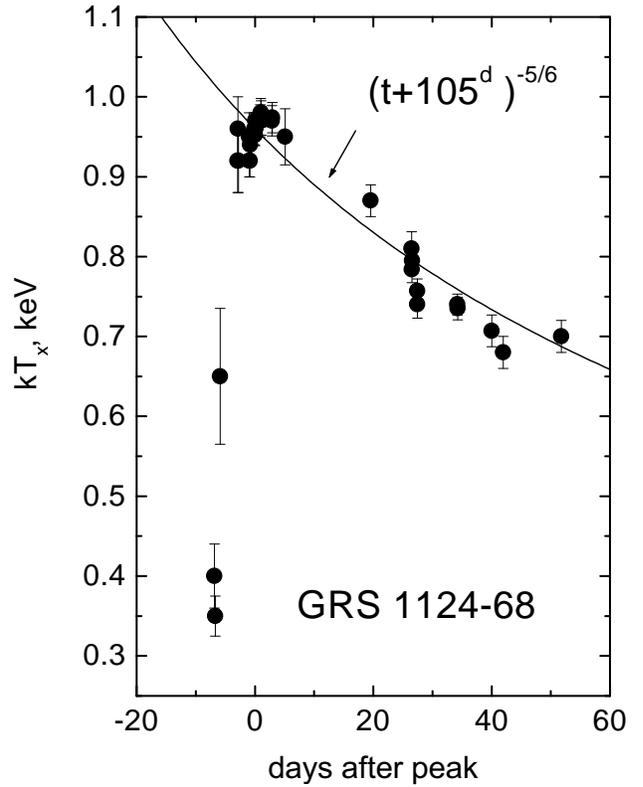}}
  \end{center}
  \caption{Observed maximum color temperature of the
disk in \mus{}~\citep{ebisawa_et1994} vs. time in comparison with
Eq.~\eqref{est.tx} where $\delta t = 105$~days expressed in
seconds.
  \label{fig.T_t}
  }
\end{figure}

An estimate of $R_\mathrm{out}$ can be derived from
Eqs.~\eqref{semiaxis} and \eqref{rout} in the approximation that
$M_\mathrm{bh} \gg M_\mathrm{o}$
\be
R_\mathrm{out} \propto M_\mathrm{bh}^{1/3}\, P^{2/3}\, .
\ee
Using this relation and Eq.~(\ref{Mdot_sol}), we obtain
\be
\dot{M} \propto \frac{M_\mathrm{bh}^{20/9}\, P^{25/9}}
                  {\alpha^{8/3}\, (t+\delta t)^{10/3}}\, .
\label{est.accrate}
\ee
Next, the bolometric luminosity of the disk can be written as
\be
L_\mathrm{bol} = \varepsilon(a)\, \dot{M} \, c^2
\approx 2\, \pi\, \sigma\, \left(\frac{T_\mathrm{X}}{f_\mathrm{c}}\right)^4\, R_\mathrm{X}^2\, ,
\label{est.lumin}
\ee
where $\varepsilon(a)$ is the efficiency of conversion of gravitational
energy to radiation, dependent on the dimensionless Kerr parameter $a$;
$T_\mathrm{X}$ is the maximum color temperature at radius
\be
R_\mathrm{X} \sim R_\mathrm{in} \propto \beta(a)\, M_\mathrm{bh}\, .
\label{est.rx}
\ee
Here $\beta(a)$ represents a dependence on the Kerr parameter. For a
Schwarzschild black hole $a=0$, $\epsilon=0.057$, $\beta=6$ and for a
maximally rotating Kerr black hole $a=1$, $\epsilon=0.42$, $\beta=1$.
Recall that the color temperature is $f_\mathrm{c}$ times greater than
the effective temperature. From Eqs.~\eqref{est.accrate} and
\eqref{est.lumin} we have
\be
T_\mathrm{X} \propto \frac{\varepsilon^{1/4}}{\beta^{1/2}} \,
\frac{M_\mathrm{bh}^{1/18}\, P^{25/36}}{\alpha^{2/3}\, (t+\delta t)^{5/6}}
\, .
\label{est.tx}
\ee
\citet{ebisawa_et1994} obtained the maximum color temperature
variations for \mus{}. Figure~\ref{fig.T_t} shows good agreement between
the observed decrease and the theoretical dependence on time
(Eq.~\ref{est.tx}).

The fading of an X-ray light curve is faster than that of a bolometric
light curve, expressed by Eq.~\eqref{est.accrate}. An interval of the
spectrum that we consider can be approximated by Wien's law with a
temperature close to the maximum color temperature of the disk. Thus,
variation in the X-ray flux of a soft X-ray transient is governed by the
decreasing color temperature and the associated shift of the Wien-like
spectrum to lower energies; a well-known exponential dependence of the
light curve on time emerges \citepalias{lip-sha2000}. The optical light
curves have flatter slopes because the main decrease in the disk
luminosity happens at high energies.

Let us calculate the slope of an optical light curve, for example, in
the $B$ band. Because the optical flux from the disk is mainly
determined by the reprocessed X-ray emission from the central source, we
can write
\be
L_B \propto R_\mathrm{out}^2 \, T_\mathrm{irr}^k\, ,
\ee
where index $k\geqslant1$ comes from a relation between the optical flux
and the effective temperature. For instance, in the Rayleigh-Jeans part
of the spectrum $k=1$. The time dependence of the effective temperature
at the outer edge of the disk can be obtained from Eq.~\eqref{eq.F_irr}:
\be
\sigma T_{\rm irr}^4 \propto \eta\, \frac{L_{\rm bol}}{R_{\rm out}^2} \left(\frac{z_0}{R_{\rm
out}}\right)^2\,
\ee
or, using Eqs.~\eqref{eq.z/r}, \eqref{est.accrate}, and
\eqref{est.lumin},
\be
T_{\rm irr} \propto (\varepsilon\, \eta)^{1/4}\,
 \frac{M_\mathrm{bh}^{7/18}\, P^{11/18}}{\alpha^{11/12}\,
(t+\delta t)^{13/12}}\, .
\label{est.tirr}
\ee

The slope of the $B$ light curve at time $t$  is 
\be
a_B = \frac{\mathrm{d}\lg L_B(t) }{\mathrm{d} t} =
-\frac{13}{12}\,\frac{k\, \lg{\mathrm{e}}}{t+\delta t}\, ,
\label{est.slope}
\ee
and is essentially determined only by the parameter $\delta t$ and vice
versa. Our numerical estimations yield $k\approx 1.9$ for the spectrum
of an irradiated disk.
Substituting values of slopes $a_B$ from Table~\ref{t.obs_data} into
Eq.~(\ref{est.slope}), we obtain that $\delta t$ is approximately 90 and
100--110 days for \mon{} and \mus{} (for the 30th and the 50th days after
the peak, respectively).
\begin{table}
\begin{minipage}[t]{\columnwidth}
\caption{Observed input parameters and fitted
light curve parameters$^a$.}
\label{t.obs_data}
\centering
\renewcommand{\footnoterule}{}
\begin{tabular}{lll}
\hline\hline
Parameter      & \mon                    &  \mus                 \\
\hline
$P$ [days]        & $0.322^{1,2}$ & $0.433^5$  \\
$f(M)$           & $2.7 M_{\sun}^{1,2}$
                                           & $3.0 M_{\sun}^5$\\
$M_\mathrm{o}$   & $0.5 M_{\sun}$          & $0.8 M_{\sun} $      \\
$N_\mathrm{H}$ [atoms cm$^{-2}$]
                 & $3 \times 10^{21}\,^{\mathrm{b,c}}$
                                           & $1.4\times
10^{21}\,^{\mathrm{c},6}$\\
$E(B-V)$         & $0.35^3$     & $0.3^{\mathrm{c}}$   \\
$\ax$            & $-0.0149 \pm 0.0004$     & $-0.0125\pm 0.0005$     \\
$\Fx$            & $100$~photons~s$^{-1}$~cm$^{-2}$ & $6.8\times
					10^{-8}$~erg~s$^{-1}$~cm$^{-2}$    \\
$a_{B}$          & $-0.0074\pm 0.0003$     & $-0.0060\pm 0.0004$   \\
$F_B$ [erg~s~cm$^2$]
                 & $ (1.24\pm 0.07)\times 10^{-10}$
		                           & $(1.12\pm0.03)\times 10^{-11}$\\
$B-V$            & $0\fm24 \pm 0\fm03^4$
                                           & $0\fm27 \pm 0\fm07^4$  \\
$U-B$            & $-0\fm75 \pm 0\fm05$    & \ldots                 \\
\hline
\end{tabular}
\begin{list}{}{}
	\item[$^{\mathrm{a}}$]
Notations are explained
in Sects.~\ref{ss.modeled_p} and \ref{ss.input_p}.
Absolute fluxes are derived for the 30th day after the outburst peak.
The slopes of the X-ray light curves
and  the X-ray fluxes correspond to different X-ray intervals:
$1.5-6$~keV for \mon{} and
$1.2-6$~keV for \mus{}.
	\item[$^{\mathrm{b}}$]
Intermediate between  the values obtained by
\citet{weav-will1974} and \citet{wu_et1983}.
	\item[$^{\mathrm{c}}$]
$N_\mathrm{H}$
is obtained from $E(B-V)$ following \citet{zombeck1990}.
	\item[] References.
(1) \citet{cherep2000};
(2) \citet{chen_et1997};
(3) \citet{wu_et1983};
(4) \citet{lip-sha2002};
(5) \citet{orosz_et1996};
(6) \citet{dellav1991,cheng_et1992,shra-gonz1993}
\end{list}
\end{minipage}
\end{table}

The color index $B-V$ obviously determines the temperature at the outer
disk edge $T_\mathrm{irr}(R=R_\mathrm{out})$. Thus, the parameters $a$,
$M_\mathrm{bh}$, and $\alpha$, which produce a satisfactory model,
should be related to meet the condition
$T_\mathrm{irr}(R=R_\mathrm{out})\approx const$, defined by
Eq.~(\ref{est.tirr}).
To conclude, we note a basic feature: for a constant black hole mass and
$T_\mathrm{X}$, an increase in the Kerr parameter will cause the
distance to the binary to decrease~(cf.~Eqs.~\eqref{est.lumin} and
\eqref{est.rx}).

\section{Outbursts of Nova Mon 1975~(A\,0620$-$00)
and Nova Mus 1991~(GRS\,1124$-$68)\label{s.results}}

Observational parameters for  Nova Mon 1975 and Nova Mus 1991
are summarized in Table~\ref{t.obs_data}. We take $B$ and $V$ optical
light curves collected in \citetalias{lip-sha2002}. The $U$ light curve
of \mon{} is taken from \citet{liutyi1976}, \citet{vandenbergh1976}, and
\citet{duer-walt1976} and transferred to  units of
erg~cm$^{-2}$~s$^{-1}$ in the same way as described in
\citetalias{lip-sha2002}.

 By and large, the outbursts are very much alike. Let us point out the
distinctions. In GRS\,1124$-$68, all light curves decline more slowly:
the decay time scales, or ``$e$-folding times'', of the X-ray and
optical light curves are 35~days and 68 days for \mus{} as against
30~days and 57 days for \mon{}. In addition, the
optical-to-X-ray flux ratio at the peak in \mon{} is $\sim 2$ times higher than in
\mus. The color index $(B-V)$ is greater for \mus{} (the
object is ``redder''), although the color excess $E(B-V)$ is higher for
\mon.

The input parameters of the model are varied within
the following limits: $M_\mathrm{bh}: 5-15 M_{\sun}$,
the Kerr parameter $a$: $0-0.9981$,
the $\alpha$-parameter: $0.1-1$, the time normalization parameter
$\delta t$:  $80-140$~days. The modeling results critically
depend on the reprocessing efficiency $\eta$, which we vary between 0
and 1.

\subsection{\mon. Importance of irradiation\label{subsect.mon}}

We use the $1.5-6$~keV light curve obtained with the \hbox{SAS-3}
CSL-A~\citep{kuulkers1998}. The flux data are converted from
counts\,s$^{-1}$ to photons\,cm$^{-2}$\,s$^{-1}$ assuming a Crab-like
spectrum. The following normalizations are adopted: $1$ Crab~$\approx
136$~counts/s~\citep{buff_et1977}, $1$ Crab~$\approx
5$~photons\,cm$^{-2}$\,s$^{-1}$ in $1.5-6$~keV. We believe the resulting
error to be acceptable because $1.5-6$~keV and $3-6$~keV fluxes in Crab
units are close to each other~\citep{kaluz_et1977}, and thus the
spectral shapes of \mon{} and the Crab should be similar in
the X-ray range in question.
\begin{figure}
  \begin{center}
  \resizebox{\hsize}{!}{\includegraphics[trim= 0cm 0cm 0cm 2cm,clip]{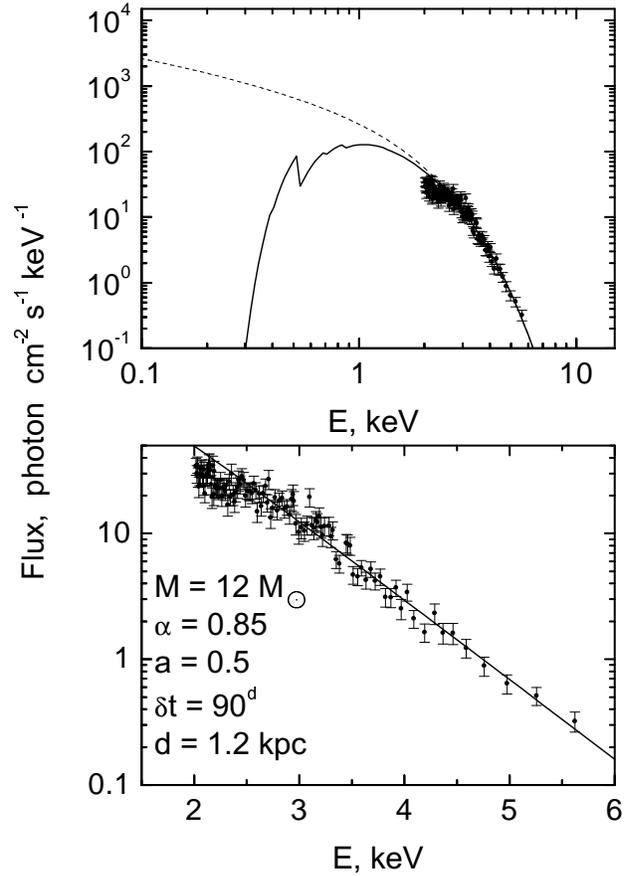}}
  \end{center}
  \caption{ Comparison of the observed
  spectrum of \mon{}~\citep{long-kest1978} and the modeled
one for $t=50^{\mathrm{d}}$.
 The solid line represents the case when interstellar absorption is
taken into account with $N_{\mathrm{H}}=3\times10^{22}$~atoms\,cm$^{-1}$;
the dashed line shows the unabsorbed modeled spectrum. The lower panel
shows a enlarged area from the upper panel. \label{mon.spectr.ris} }
\end{figure}
In Fig.~\ref{mon.spectr.ris} we compare the X-ray spectrum calculated
from the model with the spectrum observed by \citet{long-kest1978}. The
latter was obtained on the 70th day after the peak, when the flux had
the same magnitude as on the 50th day~\citep{kuulkers1998}. We conclude
that the chosen normalizations are satisfactory.

\begin{figure}
  \begin{center}
  \resizebox{\hsize}{!}{\includegraphics[trim= 0cm 1.4cm 0cm 2cm,clip]{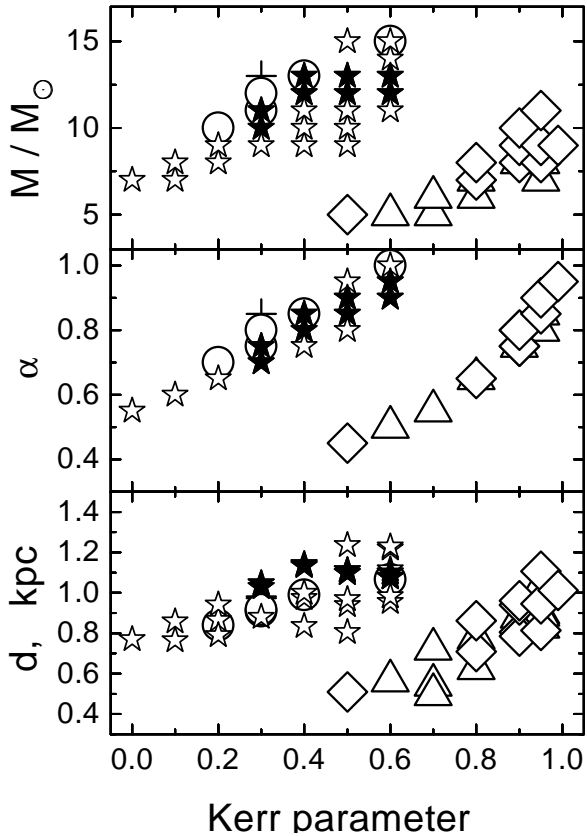}}
  \end{center}
  \caption{Resulting parameters of \mon{} vs. Kerr parameter.
Models with  $f_\mathrm{c}=1.7$ are denoted by
  circles  ($z=z_0$ and $\eta$=1),
  stars ($z=2 z_0$ and $\eta$=0.5),
 and pluses ($z=2 z_0$ and $\eta$=0.5 with limb darkening).
Models with  $f_\mathrm{c}=1$ are denoted by
  diamonds ($z=z_0$ and $\eta$=1) and
  triangles ($z=2 z_0$ and $\eta$=0.5).
Models denoted by filled stars have masses and distances in
agreement with~\citet{gelino_et2001a}.
  \label{models.mon.ris}
  }
\end{figure}

During the interval of $20-40$~days after the peak the multi-temperature
black body disk component dominates the spectrum
\citepalias[see][]{lip-sha2002}. In Table~\ref{t.obs_data} the slopes of
the light curves are calculated for this time interval, and values
$F_\mathrm{X}$ and $F_B$ correspond to the 30th day.

\begin{figure}
  \begin{center}
  \resizebox{\hsize}{!}{\includegraphics[trim= 0cm 3cm 0cm 2cm,clip]{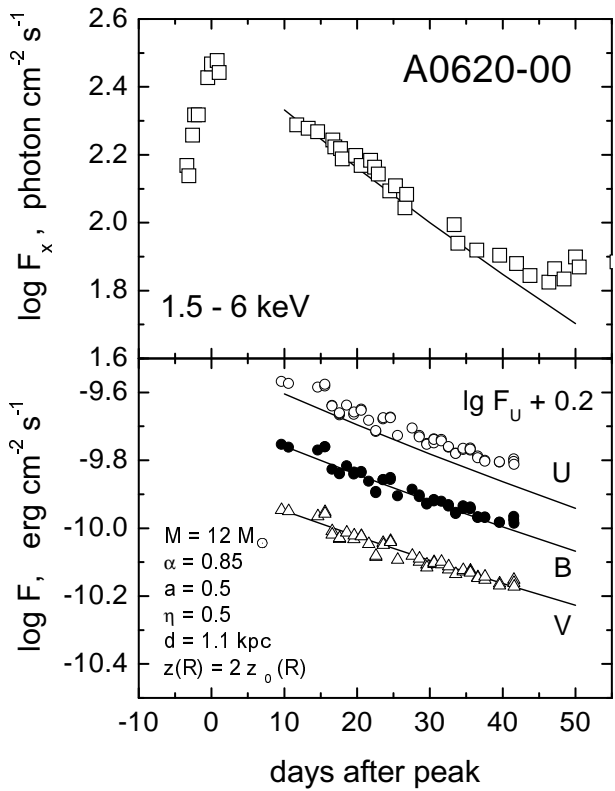}}
  \end{center}
  \caption{Comparison of the modeled and observed light curves of \mon.
  The $U$ flux is shifted upward by 0.2 .
  \label{lc.mon.ris}
  }
\end{figure}

In \citetalias{lip-sha2002} the difficulty of reproducing the slope of
the optical light curves of \mon{} was revealed. Including reprocessed
emission helps to solve the issue. In the case of the self-irradiated
disk, the slope of the optical light curve steepens, partly because the
incident flux declines faster with time than the intrinsic flux, but
mainly because $\delta t$ has to be decreased to adjust the
optical-to-X-ray flux ratio, and $\delta t$ is uniquely related to the
slope (see Eq.~\eqref{est.slope}).
Modeling results are presented in Fig.~\ref{models.mon.ris}. The
relation between allowable values of the black hole mass, the Kerr
parameter, and the $\alpha$-parameter reflects the constancy of the
value $T_{\rm X}$. An example of the modeled and observed light curves
is shown in Fig.~\ref{lc.mon.ris}.

 There are two parameters that define the effect of irradiation (see
Eq.~\eqref{eq.F_irr}): the reprocessing efficiency $\eta$ and the ratio
$z/z_0$. In the Newtonian limit the combination $\eta\, (z/z_0)^2$ is
decisive, but in general relativity this is correct only approximately
due to the relativistic function $\Psi(\theta)$. Our calculations show
that the observed parameters can be fitted with $\eta\, (z/z_0)^2 \sim
1-2$. Evidently, the reprocessing efficiency $\eta$ can hardly be as
high as 1. Therefore, the accretion disk in this X-ray nova must have an
effective geometrical thickness greater than that given by
Eq.~\eqref{eq.z/r}. One can introduce the so-called irradiation
parameter $\cal C$ to rewrite Eq.~\eqref{eq.F_irr}: $F_\mathrm{irr}(R) =
{\cal C}\, (L_\mathrm{X}/4\pi R^2)$~\citep[see, for
example,][]{dubus_et2001}.
For the model presented in Fig.~\ref{lc.mon.ris},
in the time span $10-50$~days, the
irradiation parameter, calculated at the outer radius $R_\mathrm{out} =
1.59\times10^{11}$~cm,
changes in the range $\sim (8-6)\times 10^{-4}$.

There are no satisfactory models with the Kerr parameter $a > 0.7$ for
$f_\mathrm{c}=1.7$ (Fig.~\ref{models.mon.ris}).
We also investigate the influence of parameter $f_{\rm c}$. Results with
$f_{\rm c} = 1$ are presented in Fig.~\ref{models.mon.ris} by diamonds
(for $\eta$ = 1 and $z/z_0$ = 1) and triangles ($\eta$ = 0.5 and $z/z_0$
= 2). One can see that models with $f_{\rm c} = 1$ have higher values of
the Kerr parameter, which compensate the decrease in the color
temperature. Other parameters (the black hole masses,
$\alpha$-parameter, and the distance), though somewhat decreased, are
still in the same range. Therefore, any uncertainty in $f_{\rm c}$
leads mainly to a deviation of the Kerr parameter.

In some of the models, the effect of limb darkening was included. For the
local X-ray spectra, the limb darkening was taken into account assuming
the dominant electron scattering, and the grey atmosphere (Eddington) 
approach was used  for the local optical spectra 
\citep{chandra1950B,sobolev1949,sobolev1969}. The corresponding
results are shown in Fig.~\ref{models.mon.ris} by pluses. Taking account
of limb darkening leads to some increase in the black hole mass, the
$\alpha$-parameter, and the distance to the binary, not altering the
results qualitatively.

Independent estimates of the system parameters can be used to constrain
our results further. Available in the literature are estimates of the
distance $d=1 \pm 0.4$~kpc~\citep{shahbaz_et1994,barret_et1996} and the
black hole mass: $\Mbh = 10 \pm 7 M_{\sun}$~\citep{shahbaz_et1994},
$\Mbh = 4.6 \pm 1.1 M_{\sun}$~\citep{haswell_et1993}. The highest quoted
accuracy is of the estimates given by \citet{gelino_et2001a}: $\Mbh = 11
\pm 1.9 M_{\sun}$ and $d=1.116 \pm 0.114$~kpc. The constructed models
agree with these values if $\alpha=0.7-0.95$ and the Kerr parameter
$a=0.3-0.6$ (Fig.~\ref{models.mon.ris}).

\subsection{\mus\label{ss.mus}}
\begin{figure}
 \begin{center} \resizebox{\hsize}{!}{\includegraphics[trim= 0cm 1.4cm 0cm 2cm,clip]{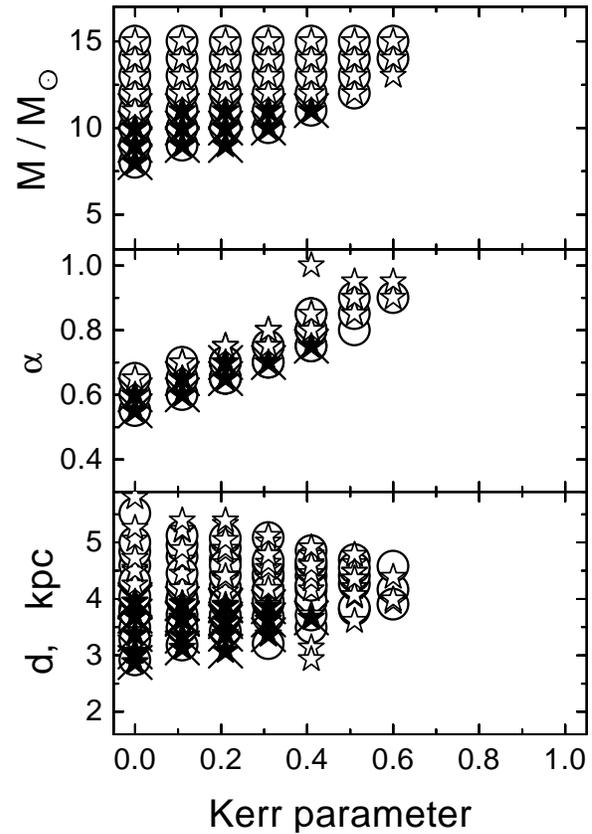}}
\end{center}
\caption{Resulting parameters of \mus{} vs. Kerr parameter. Circles
denote the models with $\eta=0.5$, $z = z_0$; stars: $\eta=0.25$, $z =
1.5 z_0$; filled stars agree with the limits on the black hole
mass~(see explanation in~\S\ref{sss.distance_mus});
crosses: models with $\eta=0.15$, $z= 2.0 z_0$ and masses within the
same limits. All models have $f_\mathrm{c}=1.7$.
\label{models.mus.ris} }
\end{figure}

\citet{ebisawa_et1994} fit the spectra obtained during the 1991 outburst
of \mus{} with a model comprising two components: a hard power-law
component and a soft component emitted by a multicolor disk
~\citep{mitsuda_et1984}. They present a time evolution of model
parameters: the maximum temperature and the inner radius of the
multicolor disk, as well as the photon index and the hard $2-20$~keV
flux from the power-law component. This information allows us to
calculate the unabsorbed $1.2-6$~keV flux. In Table~\ref{t.obs_data} the
slopes of the light curves are calculated for a 20--50 day interval.

\begin{figure}
 \begin{center} \resizebox{\hsize}{!}{\includegraphics[trim= 0cm 1.4cm 0cm 2cm,clip]{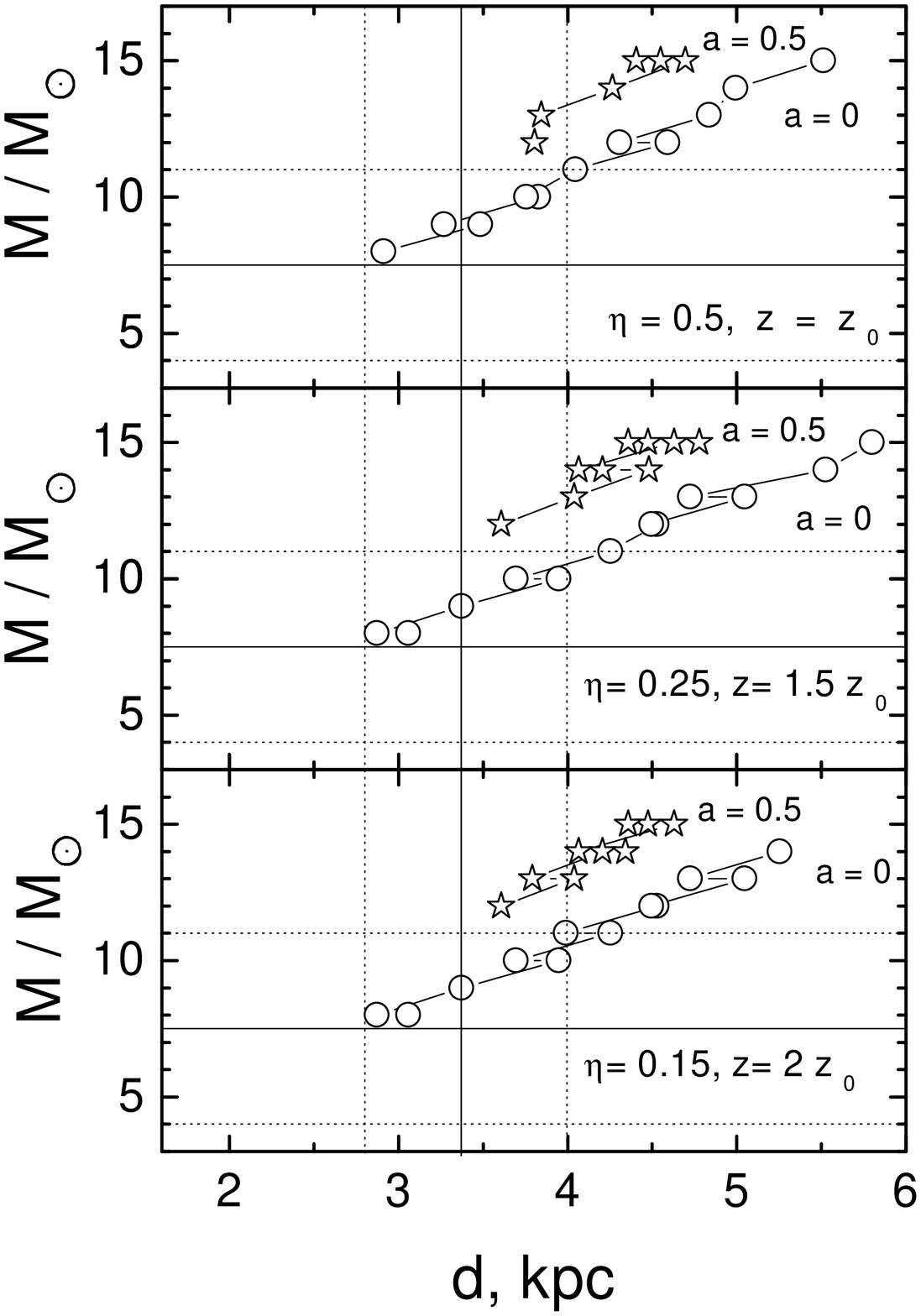}}
\end{center}
\caption{
Resulting relationships between the mass of the black hole in \mus{} and the
distance to the system, for different values of $\eta$, $z/z_0$, and the Kerr
parameter.
Horizontal and vertical lines show observational limits and central
values.
\label{fig.m-d.mus} }
\end{figure}

It is possible to fit the light curves with plausible values of
reprocessing efficiency, $\eta\lesssim0.5$, keeping the standard
thickness of the disk. For this system the relation $\eta \,(z/z_0)^2
\sim 0.5 - 0.6$ is fulfilled. One  gets almost same results with three
different sets of parameters: (1) $z=z_0$, $\eta=0.5$; (2) $z=1.5 z_0$,
$\eta=0.25$; (3) $z=2 z_0$, $\eta=0.15$ (see Figs. \ref{models.mus.ris}
and \ref{fig.m-d.mus}). The parameter $\delta t$ lies in the range from
100 to 115 days in all models. Figure~\ref{models.mus.ris} presents
allowable intervals for the mass, the Kerr parameter, the
$\alpha$-parameter, and the distance, which lies in the interval from
2.5 to 6 kpc.

\begin{figure}
  \begin{center}
  \resizebox{\hsize}{!}{\includegraphics[trim= 0cm 2.7cm 0cm 2cm,clip]{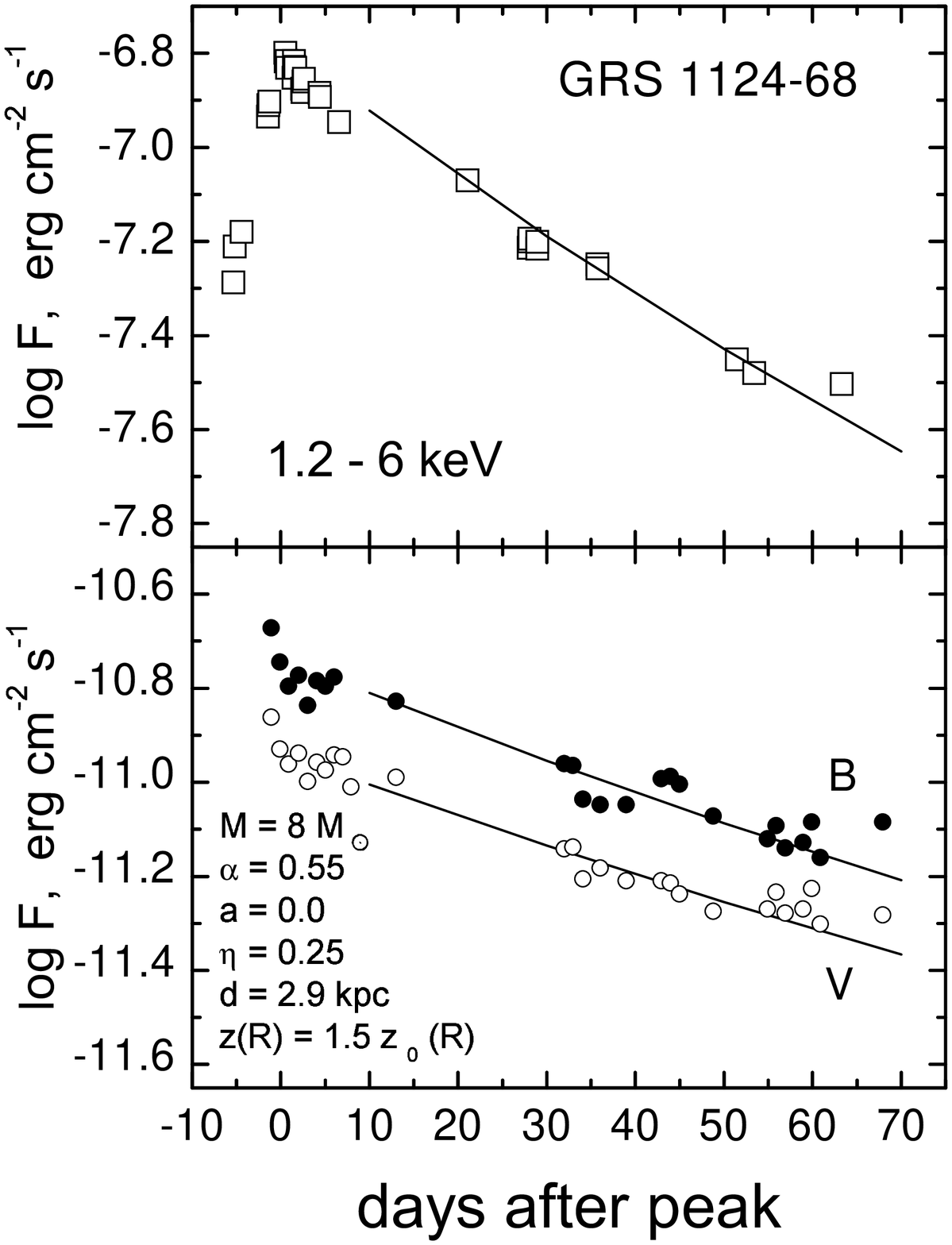}}
  \end{center}
  \caption{
  Comparison of the modeled and observed light curves of \mus.
  \label{lc.mus.ris}
  }
\end{figure}

 Figure~\ref{lc.mus.ris} illustrates the modeled light curves for
parameters $M_\mathrm{bh} = 8 M_{\sun}$, $\alpha=0.55$, $a=0$, $\delta
t=115^\mathrm{d}$, $\eta=0.25$, and $z=1.5z_0$. During the interval of
$10-60$~days, the irradiation parameter $\cal C$ (see
Sect.~\ref{subsect.mon}), calculated at the outer radius $R_\mathrm{out}
= 1.62\times10^{11}$~cm, decreases from $\sim 3\times 10^{-4}$ to
$\sim 2.5\times 10^{-4}$.

\subsubsection{Distance to \mus\label{sss.distance_mus}}

Independent estimates of the distance and the black hole mass are: 
 $d=3.4\pm 0.6$~kpc and $\Mbh = 7.5 \pm
3.5M_{\sun}$~\citep{shahbaz_et1997}, $d=5.5\pm 1$~kpc and $\Mbh = 5 -
7.5 M_{\sun}$~\citep{orosz_et1996}. The highest accuracy for the black
hole mass is claimed by~\citet{gelino_et2001b} and
\citet{khruzina_et2003e}: $\Mbh = 6.95 \pm 0.6 M_{\sun}$ and $\Mbh =
7.15 \pm 0.45 M_{\sun}$, respectively. \citet{gelino_et2001b} estimate
the distance as $5.1$~kpc or even $5.9$~kpc~\citep{gelino2004}.

The constructed models cannot agree with the distances obtained by
\citet{gelino_et2001b} and \citet{gelino2004}, tending more to the
estimates of \citet{shahbaz_et1997}. For the black hole mass $M=7.5 \pm
3.5 M_{\sun}$, $z=(1.5-2)z_0$, and $\eta=0.15- 0.25$, we come to the
following allowable intervals of parameters: the distance $3-4$~kpc, the
Kerr parameter $a\lesssim 0.4$, and the $\alpha$-parameter $0.55-
0.75$~(Fig.~\ref{models.mus.ris}).

 The above estimates of the distance are derived from an analysis of
light curves and spectra in optical and IR bands obtained during the
quiescent state of the X-ray nova. The basic element for determining the
distance in quiescence is the spectral class of the secondary. It is
found from optical spectra  \citep[$\sim$ K4V,][]{orosz_et1996} and yields
a corresponding effective temperature ($\sim 4500$~K). However, the
optical light curves of the secondary imply a reflection, and it is
necessary to assume $L_\mathrm{irr} \sim L_\mathrm{opt}$
\citep{khruzina_et2003e}. Thus, the apparent spectral class of the
secondary is biased; if the actual effective temperature of the
secondary is lower, then the binary is closer.

Another way to constrain the distance is to study X-ray spectra obtained
during the outburst. \citet{ebisawa_et1994} obtain a non-relativistic
relation between the black hole mass and the distance, which can be
rewritten as follows:
\be
d =
2.6\,\left(\frac{f_{\rm c}}{1.7}\right)^{-2}
\left(\frac{R_{\rm in}}{6R_0}\right)
\left(\frac{M_{\rm bh}}{10M_\odot}\right)
\left(\frac{\cos i}{0.5}\right)^{1/2}\rm kpc.
\ee
This relation appears to agree with our results
(Fig.~\ref{models.mus.ris}), keeping in mind that slight changes due to
relativistic corrections might be present.

It is worth noting here the possibility of increasing a distance
obtained from X-ray data. The inner disk radius can be greater than the
radius of the last marginally stable orbit $R_\mathrm{ms}$. This
phenomenon is discussed for other Galactic black hole
candidates~\citep{gilfanov_et1999,chaty_et2003}. The inner region,
evacuated by the standard disk, becomes hot and optically thin, so not
contributing to the soft X-ray flux. We tested a number of models with
the inner radius moved away up to $10 R_\mathrm{ms}$. The imposed
boundary condition was that the viscous stress tensor becomes zero at
$R_\mathrm{ms}$ and is non-zero at the inner radius $\Rin$. We came to
the conclusion that all such models, which satisfactorily fitted the
data, had higher values of the Kerr parameter and unchanged values of
the mass and the distance.

\section{Discussion and conclusions\label{s.discus}}

As shown, the model with non-stationary accretion onto a compact
binary component, developed for an $\alpha$-disk in
\citetalias{lip-sha2000}, can account for the X-ray and optical light
curves of the two X-ray novae considered, \mon{} and \mus, in the
high/soft state. Dedicated modeling of the light curves has previously
been attempted in \citetalias{lip-sha2002}. In the present study we extend
the modeling by including the effects of general relativity on light
propagation near black holes and self-irradiation of the disks.

One needs to assume a rather large contribution of reprocessed flux in
the optical light curves. A thin disk encounters difficulties here in
explaining the required amount of intercepted flux.
A twisted disk could be suggested here as a solution because
its regions facing the center undergo increased illumination~\citep{dubus_et1999,
esin_et2000},
but the total benefit, calculated for the
whole surface of one disk side, is ambiguous due to the effect of
self-shielding~\citep{suleimanov_et2007e}.

In the current study we prefer another explanation. We suggest the
presence of matter above the disk in the form of a hot corona with
temperature $\sim 2\times 10^6$~K~\citep[e.g., corresponding to the
atmosphere or the ``warm corona'' of][]{jimenez_et2002}. By intercepting
X-rays, the corona effectively increases the disk thickness. This idea
manifests itself in the computation of the optical emission when we
substitute the half-thickness of the disk $z_0$ by the value $z\approx
(1.5-2)\, z_0$ representing the height of \hbox{X-rays}' interception.
Furthermore, if the corona is inhomogeneous, i.e. consists of cooler
blobs surrounded by hotter plasma, then multiple scattering between the
blobs could explain the high values of the reprocessing
efficiency~\citep{suleim_et2003}.

The outer parts of the disks in \mon{} and \mus{} have remarkably
different properties. The disk in \mus{} turns out to be thinner and/or
and weakly thermalizes the intercepted X-ray flux. Probably the
illumination of the outer disk in \mon{} is more pronounced because of
the higher Kerr parameter. If we freeze the disk thickness at a standard
value and re-calculate models with the parameters used in
Figs.~\ref{lc.mon.ris} and \ref{lc.mus.ris}, we find that the irradiating
flux in \mon{} is twice that in \mus{}. This could be the reason for the
more developed corona above the outer disk in \mon{}.
Another possible source for the extended atmosphere of the disk is
the magnetic pressure supporting the outer layers of the disk
\citep[e.g.][]{hir2006}.  This would agree with the higher
value of $\alpha$ in \mon{}, given that the magnetorotational instability
generates the viscosity.

To bound the resulting parameters we need to invoke independent
estimates of the mass and the distance. Then for \mon{} we obtain the
intervals: $\alpha=0.7- 0.95$ and the Kerr parameter
$a=0.3-0.6$. For \mus{} we find $\alpha= 0.55 - 0.75$ and the Kerr
parameter \hbox{$a \lesssim 0.4$} if the distance to the system lies
between 3 and 4~kpc. If we do not apply the knowledge of the mass and the
distance, we are nevertheless able to confirm the high value of the
turbulent parameter: $\alpha\gtrsim 0.4$ (middle panels in
Figs.~\ref{models.mon.ris} and \ref{models.mus.ris}). Such values of
$\alpha$ are in aproximate agreement, but slightly higher than
the commonly proposed values for thin, fully ionized disks:
\hbox{$\alpha\sim 0.1 - 0.4$}~\citep[e.g.,][]{King_et2007}.

The accuracy of the estimates of $\alpha$ obtained suffers from
limitations in the calculations of the X-ray flux made in the model and
from the observational data. In this regard, we should also mention a
simplified account of the tidal effects in the analytic model of a
non-stationary disk~\citepalias{lip-sha2000}. Presumably, the accretion
disk in an X-ray nova binary has a fixed outer radius, at which a delta
function-type transport of angular momentum to the orbital motion
occurs due to tidal interactions.
A physically more comprehensive consideration, involving a real form
of the tidal torque~\citep{pap-pri1977,smak1984a,ich-osa1994,hameury-lasota2005},
could revise the solution (Eq.~\ref{Mdot_sol}) and thus change the
estimates of $\alpha$ (see Eq.~\eqref{est.accrate}). From this point of
view, the value of $\alpha$ obtained for the outer disk is a
``characteristic'' parameter of the model. However,  an
advanced treatment would not change significantly the theoretical $\dot
M$, and thus $\alpha$, if the tidal interactions are small in the
whole disk and rise very steeply only near $R_\mathrm{out}$
\citep{ich-osa1994}.

Our estimates of the Kerr parameter basically accord with a
variety of estimates recently obtained for se\-ve\-ral stellar mass
black holes: $a = 0.77\pm0.05$ for M33\,X$-$7, $a=0.65 - 0.75$ for
GRO\,J1655$-$40, $a=0.75 - 0.85$ for 4U\,1543$-$47, $a=0.98 - 1.0$ for
GRS\,1915$+$105, $a< 0.26$ for LMC\,X$-$3
\citep{Liu_et2008X,Shafee_et2006,mcclintock_et2006,Davis_et2006}. In
these works, X-ray spectra are fitted by spectral models of a
relativistic accretion disk~\citep{li_et2005,davis-hubeny2006}, and the
black holes masses and orbital inclination angles are well known.
Also, the Fe line method has been used to obtain an estimate for the
black hole spin $a = 0.86\pm0.05$ in GX\,339$-$4~\citep{Miller_et2008X}.

The model presented here can only be applied during the first few
tens of days after the peak and fails to explain the late
outburst phases.  In the low/hard state, the thin disk model cannot
adequately account for the observed spectra and energetics of X-ray novae.
The ADAF model is believed to be appropriated instead
\citep{esin_et1997,esin_et2000}. On the other hand, complex X-ray light
curves at the early outburst phases, for example those of
GRO\,J1655-40 and XTE\,J1550-564, can hardly be interpreted by means of the
present method. These binaries are highly inclined to the line of sight,
and the flat thin disks are obscured by
the outer thick disks during the early outburst \citep[see][]{nar_mcc05}.
The law of accretion rate evolution (Eq.~\ref{Mdot_sol}) also does not hold
for the events involving an additional enhanced mass transfer from the
secondary.

Our choice of the two X-ray novae stems from  their
exhibiting the most classical shape of X-ray nova light curves. The
prospect of a future determination of $\alpha$ in ionized accretion
disks in binary systems is likely to come from mo\-de\-ling the
evolution of disk parameters during the time when the disk spectral
component dominates in the spectrum of an exponentially decaying X-ray
nova in the high/soft state, and preferably with known parameters and
orbital elements of the binary.

\begin{acknowledgements} The X-ray data for \mon{}
were kindly provided by E. Kuulkers. We thank R.~Porkas, K.~Postnov,
and the anonymous referee for
useful comments. The work was supported by the Russian Foundation for
Basic Research (projects \hbox{06--02--16025} and \hbox{09--02--00032})
and by the President's Program for Support of Leading Science Schools (grant
Nsh.--4224.2008.2). V.\,F.~Suleimanov is supported by the DFG grant
SFB/Transregio 7 "Gravitational Wave Astronomy". N.\,I.~Shakura is
grateful to the Max Planck Institute for Astrophysics (Garching,
Germany) for the opportunity for annual short-term study
visits. G.\,V.~ Lipunova is grateful to the Offene
Ganztagesschule of the Paul-Klee-Grundschule (Bonn, Germany) and Stadt Bonn
for providing a possibility for her full-day scientific activity.
\end{acknowledgements}

\bibliographystyle{aa}

\begin{thebibliography}{85}
\expandafter\ifx\csname natexlab\endcsname\relax\def\natexlab#1{#1}\fi

\bibitem[{{Bardeen} \& {Petterson}(1975)}]{bp1975}
{Bardeen}, J.~M., \& {Peterson}, J.~A. 1975, \apj, 195,
L65


\bibitem[{{Barret} {et~al.}(1996){Barret}, {McClintock}, \&
  {Grindlay}}]{barret_et1996}
{Barret}, D., {McClintock}, J.~E., \& {Grindlay}, J.~E. 1996, \apj, 473, 963

\bibitem[{{Begelman} \& {McKee}(1983)}]{begel-mckee1983}
{Begelman}, M.~C. \& {McKee}, C.~F. 1983, \apj, 271, 89

\bibitem[{{Begelman} {et~al.}(1983){Begelman}, {McKee}, \&
  {Shields}}]{begelman_et1983}
{Begelman}, M.~C., {McKee}, C.~F., \& {Shields}, G.~A. 1983, \apj, 271, 70

\bibitem[{{Buff} {et~al.}(1977){Buff}, {Jernigan}, {Laufer}, {Bradt}, {Clark},
  {Lewin}, {Matilsky}, {Mayer}, \& {Primini}}]{buff_et1977}
{Buff}, J., {Jernigan}, G., {Laufer}, B., {et~al.} 1977, \apj, 212, 768

\bibitem[{{Cannizzo}(1993)}]{cannizzo1993}
{Cannizzo}, J.~K. 1993, {The Limit Cycle Instability in Dwarf Nova Accretion
  Disks} (Accretion Disks in Compact Stellar Systems), 6

\bibitem[{{Chandrasekhar}(1950)}]{chandra1950B}
{Chandrasekhar}, S. 1950, {Radiative transfer.} (Oxford, Clarendon Press,
  1950.)

\bibitem[{{Chaty} {et~al.}(2003){Chaty}, {Haswell}, {Malzac}, {Hynes},
  {Shrader}, \& {Cui}}]{chaty_et2003}
{Chaty}, S., {Haswell}, C.~A., {Malzac}, J., {et~al.} 2003, \mnras, 346, 689

\bibitem[{{Chen} {et~al.}(1997){Chen}, {Shrader}, \& {Livio}}]{chen_et1997}
{Chen}, W., {Shrader}, C.~R., \& {Livio}, M. 1997, \apj, 491, 312

\bibitem[{{Cheng} {et~al.}(1992){Cheng}, {Horne}, {Panagia}, {Shrader},
  {Gilmozzi}, {Paresce}, \& {Lund}}]{cheng_et1992}
{Cheng}, F.~H., {Horne}, K., {Panagia}, N., {et~al.} 1992, \apj, 397, 664

\bibitem[{{Cherepashchuk}(2000)}]{cherep2000}
{Cherepashchuk}, A.~M. 2000, Space Science Reviews, 93, 473

\bibitem[{{Cunningham}(1975)}]{cunnin1975}
{Cunningham}, C.~T. 1975, \apj, 202, 788

\bibitem[{{Davis} {et~al.}(2005){Davis}, {Blaes}, {Hubeny}, \&
  {Turner}}]{davis_et2005}
{Davis}, S.~W., {Blaes}, O.~M., {Hubeny}, I., \& {Turner}, N.~J. 2005, \apj,
  621, 372

\bibitem[{{Davis} {et~al.}(2006){Davis}, {Done}, \& {Blaes}}]{Davis_et2006}
{Davis}, S.~W., {Done}, C., \& {Blaes}, O.~M. 2006, \apj, 647, 525

\bibitem[{{Davis} \& {Hubeny}(2006)}]{davis-hubeny2006}
{Davis}, S.~W. \& {Hubeny}, I. 2006, \apjs, 164, 530

\bibitem[{{de Jong} {et~al.}(1996){de Jong}, {van Paradijs}, \&
  {Augusteijn}}]{dejong_et1996}
{de Jong}, J.~A., {van Paradijs}, J., \& {Augusteijn}, T. 1996, \aap, 314, 484

\bibitem[{{della Valle} {et~al.}(1991){della Valle}, {Jarvis}, \&
  {West}}]{dellav1991}
{della Valle}, M., {Jarvis}, B.~J., \& {West}, R.~M. 1991, \nat, 353, 50

\bibitem[{{Dubus} {et~al.}(2001){Dubus}, {Hameury}, \& {Lasota}}]{dubus_et2001}
{Dubus}, G., {Hameury}, J.-M., \& {Lasota}, J.-P. 2001, \aap, 373, 251

\bibitem[{{Dubus} {et~al.}(1999){Dubus}, {Lasota}, {Hameury}, \&
  {Charles}}]{dubus_et1999}
{Dubus}, G., {Lasota}, J.-P., {Hameury}, J.-M., \& {Charles}, P. 1999, \mnras,
  303, 139

\bibitem[{{Duerbeck} \& {Walter}(1976)}]{duer-walt1976}
{Duerbeck}, H.~W. \& {Walter}, K. 1976, \aap, 48, 141

\bibitem[{{Ebisawa} {et~al.}(1994){Ebisawa}, {Ogawa}, {Aoki}, {Dotani},
  {Takizawa}, {Tanaka}, {Yoshida}, {Miyamoto}, {Iga}, {Hayashida}, {Kitamoto},
  \& {Terada}}]{ebisawa_et1994}
{Ebisawa}, K., {Ogawa}, M., {Aoki}, T., {et~al.} 1994, \pasj, 46, 375

\bibitem[{{Eggleton}(1983)}]{egglton1983}
{Eggleton}, P.~P. 1983, \apj, 268, 368

\bibitem[{{El-Khoury} \& {Wickramasinghe}(1999)}]{elk-wic1999}
{El-Khoury}, W. \& {Wickramasinghe}, D. 1999, \mnras, 303, 380

\bibitem[{{Esin} {et~al.}(2000){Esin}, {Kuulkers}, {McClintock}, \&
  {Narayan}}]{esin_et2000}
{Esin}, A.~A., {Kuulkers}, E., {McClintock}, J.~E., \& {Narayan}, R. 2000,
  \apj, 532, 1069

\bibitem[{{Esin} {et~al.}(1997){Esin}, {McClintock}, \&
  {Narayan}}]{esin_et1997}
{Esin}, A.~A., {McClintock}, J.~E., \& {Narayan}, R. 1997, \apj, 489, 865

\bibitem[{{Frank} {et~al.}(2002){Frank}, {King}, \& {Raine}}]{fkr_book2002}
{Frank}, J., {King}, A., \& {Raine}, D.~J. 2002, {Accretion Power in
  Astrophysics}, 3rd edn. (Cambridge, UK: Cambridge University Press)

\bibitem[{{Gelino}(2004)}]{gelino2004}
{Gelino}, D.~M. 2004, in Revista Mexicana de Astronomia y Astrofisica
  Conference Series, Vol.~20, Revista Mexicana de Astronomia y Astrofisica
  Conference Series, ed. G.~{Tovmassian} \& E.~{Sion}, 214--214

\bibitem[{{Gelino} {et~al.}(2001{\natexlab{a}}){Gelino}, {Harrison}, \&
  {McNamara}}]{gelino_et2001a}
{Gelino}, D.~M., {Harrison}, T.~E., \& {McNamara}, B.~J. 2001{\natexlab{a}},
  \aj, 122, 971

\bibitem[{{Gelino} {et~al.}(2001{\natexlab{b}}){Gelino}, {Harrison}, \&
  {Orosz}}]{gelino_et2001b}
{Gelino}, D.~M., {Harrison}, T.~E., \& {Orosz}, J.~A. 2001{\natexlab{b}}, \aj,
  122, 2668

\bibitem[{{Gilfanov} {et~al.}(1999){Gilfanov}, {Churazov}, \&
  {Revnivtsev}}]{gilfanov_et1999}
{Gilfanov}, M., {Churazov}, E., \& {Revnivtsev}, M. 1999, \aap, 352, 182

\bibitem[{{Hameury} \& {Lasota}(2005)}]{hameury-lasota2005}
{Hameury}, J.-M. \& {Lasota}, J.-P. 2005, \aap, 443, 283

\bibitem[{{Haswell} {et~al.}(1993){Haswell}, {Robinson}, {Horne}, {Stiening},
  \& {Abbott}}]{haswell_et1993}
{Haswell}, C.~A., {Robinson}, E.~L., {Horne}, K., {Stiening}, R.~F., \&
  {Abbott}, T.~M.~C. 1993, \apj, 411, 802

\bibitem[{{Hirose} {et~al.}(2006)}]{hir2006}
{Hirose}, S., {Krolik}, J.~H., \& {Stone}, J.~M. 2006, \apj, 640, 901

\bibitem[{{Ichikawa} \& {Osaki}(1994)}]{ich-osa1994}
{Ichikawa}, S. \& {Osaki}, Y. 1994, \pasj, 46, 621

\bibitem[{{Jimenez-Garate} {et~al.}(2002){Jimenez-Garate}, {Raymond}, \&
  {Liedahl}}]{jimenez_et2002}
{Jimenez-Garate}, M.~A., {Raymond}, J.~C., \& {Liedahl}, D.~A. 2002, \apj, 581,
  1297

\bibitem[{{Kaluzienski} {et~al.}(1977){Kaluzienski}, {Holt}, {Boldt}, \&
  {Serlemitsos}}]{kaluz_et1977}
{Kaluzienski}, L.~J., {Holt}, S.~S., {Boldt}, E.~A., \& {Serlemitsos}, P.~J.
  1977, \apj, 212, 203

\bibitem[{{Ketsaris} \& {Shakura}(1998)}]{ket-sha1998}
{Ketsaris}, N.~A. \& {Shakura}, N.~I. 1998, Astronomical and Astrophysical
  Transactions, 15, 193

\bibitem[{{Khruzina} {et~al.}(2003){Khruzina}, {Cherepashchuk}, {Bisikalo},
  {Boyarchuk}, \& {Kuznetsov}}]{khruzina_et2003e}
{Khruzina}, T.~S., {Cherepashchuk}, A.~M., {Bisikalo}, D.~V., {Boyarchuk},
  A.~A., \& {Kuznetsov}, O.~A. 2003, Astronomy Reports, 47, 621

\bibitem[{{King} {et~al.}(2007){King}, {Pringle}, \& {Livio}}]{King_et2007}
{King}, A.~R., {Pringle}, J.~E., \& {Livio}, M. 2007, \mnras, 376, 1740

\bibitem[{{King} \& {Ritter}(1998)}]{kin-rit1998}
{King}, A.~R. \& {Ritter}, H. 1998, \mnras, 293, L42

\bibitem[{{Ko} \& {Kallman}(1994)}]{ko-kallman1994}
{Ko}, Y.-K. \& {Kallman}, T.~R. 1994, \apj, 431, 273

\bibitem[{{Kurucz}(1994)}]{kurucz1994_n19-22}
{Kurucz}, R. 1994, CD-ROMs No.~19-22.~ Cambridge, Mass.: Smithsonian
  Astrophysical Observatory, 1994.

\bibitem[{{Kuulkers}(1998)}]{kuulkers1998}
{Kuulkers}, E. 1998, New Astronomy Review, 42, 1

\bibitem[{{Lasota}(2001)}]{lasota2001}
{Lasota}, J.-P. 2001, New Astronomy Review, 45, 449

\bibitem[{{Li} {et~al.}(2005){Li}, {Zimmerman}, {Narayan}, \&
  {McClintock}}]{li_et2005}
{Li}, L.-X., {Zimmerman}, E.~R., {Narayan}, R., \& {McClintock}, J.~E. 2005,
  \apjs, 157, 335

\bibitem[{{Lipunova} \& {Shakura}(2000)}]{lip-sha2000}
{Lipunova}, G.~V. \& {Shakura}, N.~I. 2000, \aap, 356, 363

\bibitem[{{Lipunova} \& {Shakura}(2002)}]{lip-sha2002}
{Lipunova}, G.~V. \& {Shakura}, N.~I. 2002, Astronomy Reports, 46, 366

\bibitem[{{Lipunova} \& {Suleimanov}(2004)}]{lip-sul2004}
{Lipunova}, G.~V. \& {Suleimanov}, V.~F. 2004, Baltic Astronomy, 13, 167

\bibitem[{{Liu} {et~al.}(2008){Liu}, {McClintock}, {Narayan}, {Davis}, \&
  {Orosz}}]{Liu_et2008X}
{Liu}, J., {McClintock}, J., {Narayan}, R., {Davis}, S., \& {Orosz}, J. 2008,
  ArXiv e-prints, 803

\bibitem[{{Liutyi}(1976)}]{liutyi1976}
{Liutyi}, V.~M. 1976, Soviet Astronomy Letters, 2, 43

\bibitem[{{Long} \& {Kestenbaum}(1978)}]{long-kest1978}
{Long}, K.~S. \& {Kestenbaum}, H.~L. 1978, \apj, 226, 271

\bibitem[{{Lyubarskij} \& {Shakura}(1987)}]{lyub-shak1987}
{Lyubarskij}, Y.~E. \& {Shakura}, N.~I. 1987, Soviet Astronomy Letters, 13, 386

\bibitem[{{McClintock} \& {Remillard}(2003)}]{mcc-rem2003}
{McClintock}, J.~E. \& {Remillard}, R.~A. 2003, ArXiv Astrophysics e-prints:
  astro-ph/0306213

\bibitem[{{McClintock} {et~al.}(2006){McClintock}, {Shafee}, {Narayan},
  {Remillard}, {Davis}, \& {Li}}]{mcclintock_et2006}
{McClintock}, J.~E., {Shafee}, R., {Narayan}, R., {et~al.} 2006, \apj, 652, 518

\bibitem[{{Meyer} \& {Meyer-Hofmeister}(1981)}]{meye-meye1981}
{Meyer}, F. \& {Meyer-Hofmeister}, E. 1981, \aap, 104, L10

\bibitem[{{Miller} \& {Stone}(2000)}]{Miller_St2000} {Miller}, K.~J.,
{Stone}, J.~M. 2008, \apj, 534, 398

\bibitem[{{Miller} {et~al.}(2008){Miller}, {Reynolds}, {Fabian}, {Cackett},
  {Miniutti}, {Raymond}, {Steeghs}, {Reis}, \& {Homan}}]{Miller_et2008X}
{Miller}, J.~M., {Reynolds}, C.~S., {Fabian}, A.~C., {et~al.} 2008, ArXiv
  e-prints, 802

\bibitem[{{Mitsuda} {et~al.}(1984){Mitsuda}, {Inoue}, {Koyama}, {Makishima},
  {Matsuoka}, {Ogawara}, {Suzuki}, {Tanaka}, {Shibazaki}, \&
  {Hirano}}]{mitsuda_et1984}
{Mitsuda}, K., {Inoue}, H., {Koyama}, K., {et~al.} 1984, \pasj, 36, 741

\bibitem[{{Morrison} \& {McCammon}(1983)}]{mor-mcc1983}
{Morrison}, R. \& {McCammon}, D. 1983, \apj, 270, 119

\bibitem[{{Narayan} {et~al.}(1998){Narayan}, {Mahadevan}, \&
  {Quataert}}]{narayan_et1998}
{Narayan}, R., {Mahadevan}, R., \& {Quataert}, E. 1998, in Theory of Black Hole
  Accretion Disks, 148

\bibitem[{{Narayan} \& {McClintock}(2005)}]{nar_mcc05}
{Narayan}, R., \& {McClintock}, J.~F. 2005, \apj, 623, 1017


\bibitem[{{Orosz} {et~al.}(1996){Orosz}, {Bailyn}, {McClintock}, \&
  {Remillard}}]{orosz_et1996}
{Orosz}, J.~A., {Bailyn}, C.~D., {McClintock}, J.~E., \& {Remillard}, R.~A.
  1996, \apj, 468, 380

\bibitem[{{Paczynski}(1977)}]{paczynski1977}
{Paczynski}, B. 1977, \apj, 216, 822

\bibitem[{{Page} \& {Thorne}(1974)}]{page-thor1974}
{Page}, D.~N. \& {Thorne}, K.~S. 1974, \apj, 191, 499

\bibitem[{{Papaloizou} \& {Pringle}(1977)}]{pap-pri1977}
{Papaloizou}, J. \& {Pringle}, J.~E. 1977, \mnras, 181, 441

\bibitem[{{Proga} \& {Kallman}(2002)}]{proga-kallman2002}
{Proga}, D. \& {Kallman}, T.~R. 2002, \apj, 565, 455

\bibitem[{{Riffert} \& {Herold}(1995)}]{rif-her1995}
{Riffert}, H. \& {Herold}, H. 1995, \apj, 450, 508

\bibitem[{{R{\'o}{\.z}a{\'n}ska} \& {Czerny}(1996)}]{rozan-czern1996}
{R{\'o}{\.z}a{\'n}ska}, A. \& {Czerny}, B. 1996, Acta Astronomica, 46, 233

\bibitem[{{Shafee} {et~al.}(2006){Shafee}, {McClintock}, {Narayan}, {Davis},
  {Li}, \& {Remillard}}]{Shafee_et2006}
{Shafee}, R., {McClintock}, J.~E., {Narayan}, R., {et~al.} 2006, \apjl, 636,
  L113

\bibitem[{{Shahbaz} {et~al.}(1994){Shahbaz}, {Naylor}, \&
  {Charles}}]{shahbaz_et1994}
{Shahbaz}, T., {Naylor}, T., \& {Charles}, P.~A. 1994, \mnras, 268, 756

\bibitem[{{Shahbaz} {et~al.}(1997){Shahbaz}, {Naylor}, \&
  {Charles}}]{shahbaz_et1997}
{Shahbaz}, T., {Naylor}, T., \& {Charles}, P.~A. 1997, \mnras, 285, 607

\bibitem[{{Shakura} \& {Sunyaev}(1973)}]{sha-sun1973}
{Shakura}, N.~I. \& {Sunyaev}, R.~A. 1973, \aap, 24, 337

\bibitem[{{Shimura} \& {Takahara}(1995)}]{shi-tak1995}
{Shimura}, T. \& {Takahara}, F. 1995, \apj, 445, 780

\bibitem[{{Shrader} \& {Gonzalez-Riestra}(1993)}]{shra-gonz1993}
{Shrader}, C.~R. \& {Gonzalez-Riestra}, R. 1993, \aap, 276, 373

\bibitem[{{Smak}(1984)}]{smak1984a}
{Smak}, J. 1984, Acta Astronomica, 34, 161

\bibitem[{{Sobolev}(1949)}]{sobolev1949}
{Sobolev}, V.~V. 1949, Uchenye Zapiski Leningrad. Univ., Seria Matem. Nauk, 18,
  N 1163

\bibitem[{{Sobolev}(1969)}]{sobolev1969}
{Sobolev}, V.~V. 1969, 1, Vol. F-531, {Course in Theoretical Astrophysics}
  (NASA)

\bibitem[{{Speith} {et~al.}(1995){Speith}, {Riffert}, \&
  {Ruder}}]{speith_et1995}
{Speith}, R., {Riffert}, H., \& {Ruder}, H. 1995, Computer Physics
  Communications, 88, 109

\bibitem[{{Suleimanov} {et~al.}(1999){Suleimanov}, {Meyer}, \&
  {Meyer-Hofmeister}}]{suleim_et1999}
{Suleimanov}, V., {Meyer}, F., \& {Meyer-Hofmeister}, E. 1999, \aap, 350, 63

\bibitem[{{Suleimanov} {et~al.}(2003){Suleimanov}, {Meyer}, \&
  {Meyer-Hofmeister}}]{suleim_et2003}
{Suleimanov}, V., {Meyer}, F., \& {Meyer-Hofmeister}, E. 2003, \aap, 401, 1009

\bibitem[{{Suleimanov}(1991)}]{suleim1991}
{Suleimanov}, V.~F. 1991, Soviet Astronomy Letters, 17, 245

\bibitem[{{Suleimanov} {et~al.}(2007){Suleimanov}, {Lipunova}, \&
  {Shakura}}]{suleimanov_et2007e}
{Suleimanov}, V.~F., {Lipunova}, G.~V., \& {Shakura}, N.~I. 2007, Astronomy
  Reports, 51, 549

\bibitem[{{Sulkanen} {et~al.}(1981){Sulkanen}, {Brasure}, \&
  {Patterson}}]{sulkanen_et1981}
{Sulkanen}, M.~E., {Brasure}, L.~W., \& {Patterson}, J. 1981, \apj, 244, 579

\bibitem[{{Tanaka} \& {Shibazaki}(1996)}]{tan-shi1996}
{Tanaka}, Y. \& {Shibazaki}, N. 1996, \araa, 34, 607

\bibitem[{{Turner}(2004)}]{tur_04}
{Turner}, N.~J. 2004, \apj, 605, L45


\bibitem[{{van den Bergh}(1976)}]{vandenbergh1976}
{van den Bergh}, S. 1976, \aj, 81, 104

\bibitem[{{van Paradijs}(1996)}]{vanparadijs1996}
{van Paradijs}, J. 1996, \apjl, 464, L139

\bibitem[{{Weaver} \& {Williams}(1974)}]{weav-will1974}
{Weaver}, H. \& {Williams}, D.~R.~W. 1974, \aaps, 17, 1

\bibitem[{{Wu} {et~al.}(1983){Wu}, {Panek}, {Holm}, {Schmitz}, \&
  {Swank}}]{wu_et1983}
{Wu}, C.-C., {Panek}, R.~J., {Holm}, A.~V., {Schmitz}, M., \& {Swank}, J.~H.
  1983, \pasp, 95, 391

\bibitem[{{Zombeck}(1990)}]{zombeck1990}
{Zombeck}, M.~V. 1990, {Handbook of space astronomy and astrophysics}
  (Cambridge: University Press, 1990, 2nd ed.)

\end{thebibliography}


\end{document}